%% file: main.tex
\documentclass[lettersize,journal]{IEEEtran}
\usepackage[T1]{fontenc}
\usepackage{amssymb}
\usepackage[pdftex]{graphicx}
\usepackage{amsmath}
\usepackage{amsthm}
\usepackage{mathtools}
\usepackage{array}
\usepackage[noadjust]{cite}
\usepackage{algorithm}
\usepackage{algpseudocode}
\usepackage{multirow}
\usepackage{xcolor}
\usepackage[utf8]{inputenc}
\usepackage{textcomp}
\usepackage[hidelinks]{hyperref}
\usepackage[acronym]{glossaries}
\usepackage{siunitx}

\algnewcommand{\Initialize}[1]{%
  \State \textbf{Initialize}
  \Statex \hspace*{\algorithmicindent}\parbox[t]{.8\linewidth}{\raggedright #1}
}

\algnewcommand{\algorithmicgoto}{\textbf{go to}}%
\algnewcommand{\Goto}[1]{\algorithmicgoto~\ref{#1}}%

\makeatletter
\let\MYcaption\@makecaption
\makeatother

\algnewcommand{\LineComment}[1]{\(\triangleright\) #1}

\DeclareSIUnit\frame{frame}
\DeclareSIUnit{\kmph}{kmph}
\DeclareSIUnit{\dBm}{dBm}
\DeclareSIUnit{\microsecond}{\micro\second} 
\DeclareSIUnit\PRB{PRB}
\DeclareMathAlphabet{\mathbcal}{OMS}{cmsy}{b}{n}

\input{tex/abbreviations}

\input{tex/macros}

\begin{document}

\title{Joint Resource-Power Allocation and UE Rank Selection in Multi-User MIMO Systems with Linear Transceivers}
\author{K. Pavan Srinath, \textit{Member, IEEE}, Alix Jeannerot, and Alvaro Valcarce Rial, \textit{Senior Member, IEEE}
\thanks{K. P. Srinath and A. Valcarce Rial are with Nokia Bell Labs, 91300 MASSY, France (email: \{pavan.koteshwar\_srinath, alvaro.valcarce\_rial\}@nokia-bell-labs.com), and A. Jeannerot is at INSA Lyon, Inria, CITI, UR3720, 69621 Villeurbanne, France (email: alix.jeannerot@inria.fr). A significant part of this work was done when A. Jeannerot was at Nokia Bell Labs, 91300 MASSY, France.}}
\date{April 2024}

\maketitle

\begin{abstract}
\input{tex/abstract}
\end{abstract}

\glsresetall
\input{tex/introduction}
\input{tex/system_model}
\input{tex/joint_resource_allocation}
\input{tex/proposed_technique}
\input{tex/simulation_results}
\input{tex/concluding_remarks}

\bibliography{references}
\bibliographystyle{IEEEtran}

\end{document}

%% file: tex/abbreviations.tex
\newacronym{3GPP}{3GPP}{3rd Generation Partnership Project}
\newacronym{5GNR}{5G NR}{5G New Radio}

\newacronym{AM}{AM}{arithmetic mean}
\newacronym{AMP}{AMP}{approximate message passing}
\newacronym{AWGN}{AWGN}{additive white Gaussian noise}

\newacronym{BER}{BER}{bit error rate}
\newacronym{BICM}{BICM}{bit-interleaved coded modulation}
\newacronym{BLER}{BLER}{block error rate}
\newacronym{BMDR}{BMDR}{bit-metric decoding rate}
\newacronym{BS}{BS}{base station}

\newacronym{CDF}{CDF}{cumulative distribution function}
\newacronym{CER}{CER}{codeword error rate}
\newacronym{CNN}{CNN}{convolutional neural network}
\newacronym{Conv2D}{Conv2D}{convolutional}
\newacronym{CQI}{CQI}{Channel Quality Information}
\newacronym{CSI}{CSI}{channel-state information}
\newacronym{CSI-RS}{CSI-RS}{channel-state information reference signal}

\newacronym{DMRS}{DMRS}{demodulation reference signal}
\newacronym{DCI}{DCI}{downlink control information}
\newacronym{DL}{DL}{downlink}

\newacronym{eMBB}{eMBB}{enhanced mobile broadband}
\newacronym{EP}{EP}{expectation propagation}
\newacronym{EVD}{EVD}{eigenvalue decomposition}

\newacronym{FC}{FC}{fully-connected}
\newacronym{FDD}{FDD}{frequency division duplex}

\newacronym{GM}{GM}{geometric mean}
\newacronym{GMI}{GMI}{generalized mutual information}
\newacronym{gNB}{gNB}{next generation NodeB}

\newacronym{HARQ}{HARQ}{hybrid automatic repeat request}

\newacronym{iid}{i.i.d.\@}{independent and identically distributed}
\newacronym{ISI}{ISI}{inter-symbol interference}
\newacronym{ICI}{ICI}{inter-cell interference}
\newacronym{IPOPT}{IPOPT}{Interior Point OPTimizer}

\newacronym{KL}{KL}{Kullback–Leibler}

\newacronym{LA}{LA}{link-adaptation}
\newacronym{LDPC}{LDPC}{low-density parity-check}
\newacronym{LMMSE}{LMMSE}{linear minimum mean square error}
\newacronym{LLR}{LLR}{log-likelihood ratio}

\newacronym{MAP}{MAP}{maximum a posteriori}
\newacronym{MCS}{MCS}{modulation and coding scheme}
\newacronym{MI}{MI}{mutual-information}
\newacronym{MIESM}{MIESM}{mutual-information effective SINR metric}
\newacronym{MIMO}{MIMO}{multiple-input multiple-output}
\newacronym{ML}{ML}{maximum-likelihood}
\newacronym{MMSE}{MMSE}{minimum mean square error}
\newacronym{MSE}{MSE}{mean squared error}
\newacronym{MU-MIMO}{MU-MIMO}{multi-user multiple-input multiple-output}

\newacronym{NN}{NN}{neural network}

\newacronym{OFDM}{OFDM}{orthogonal frequency-division multiplexing}
\newacronym{OFDMA}{OFDMA}{orthogonal frequency-division multiple access}
\newacronym{OLLA}{OLLA}{open loop link-adaptation}
\newacronym{OLPC}{OLPC}{open loop power control}

\newacronym{PAPC}{PAPC}{per-antenna power constraint}
\newacronym{PDSCH}{PDSCH}{physical downlink shared channel}
\newacronym{PHY}{PHY}{physical layer}
\newacronym{PRB}{PRB}{physical resource block}
\newacronym{PUSCH}{PUSCH}{physical uplink shared channel}

\newacronym{QAM}{QAM}{quadrature amplitude modulation}
\newacronym{QoS}{QoS}{quality-of-service}
\newacronym{QPSK}{QPSK}{quadrature phase-shift keying}

\newacronym{RBG}{PRBG}{physical resource block group}
\newacronym{RE}{RE}{resource element}
\newacronym{RI}{RI}{rank indicator}
\newacronym{RL}{RL}{reinforcement-learning}
\newacronym{RSRP}{RSRP}{reference signal received power}

\newacronym{SD}{SD}{standard deviation}
\newacronym{SE}{SE}{spectral efficiency}
\newacronym{SGD}{SGD}{stochastic gradient descent}
\newacronym{SIC}{SIC}{soft interference cancellation}
\newacronym{SINR}{SINR}{signal-to-interference-noise ratio}
\newacronym{SISO}{SISO}{single-input single-ouput}
\newacronym{SLS}{SLS}{system-level simulations}
\newacronym{SNR}{SNR}{signal-to-noise ratio}
\newacronym{SRS}{SRS}{sounding reference signal}
\newacronym{SU-MIMO}{SU-MIMO}{single-user MIMO}
\newacronym{SU-SISO}{SU-SISO}{single-user SISO}
\newacronym{SVD}{SVD}{singular value decomposition}

\newacronym{TDD}{TDD}{time division duplex}
\newacronym{TPC}{TPC}{transmit power control}

\newacronym[longplural={user-equipment}]{UE}{UE}{user-equipment}
\newacronym{UL}{UL}{uplink}
\newacronym{uRLLC}{uRLLC}{ultra-reliable low-latency communications}

\newacronym{wrt}{w.r.t.\@}{with respect to}

\newacronym{ZF}{ZF}{zero-forcing}

%% file: tex/macros.tex
\DeclareMathAlphabet{\mathbcal}{OMS}{cmsy}{b}{n}
\renewcommand{\vec}[1]{\mathbf{#1}}

\algblock{Input}{EndInput}
\algnotext{EndInput}
\algblock{Output}{EndOutput}
\algnotext{EndOutput}
\newcommand{\Desc}[2]{\State \makebox[3em][l]{#1}#2}

\newcommand{\be}{\begin{equation}}
\newcommand{\ee}{\end{equation}}			
\newcommand{\ben}{\begin{equation*}}
\newcommand{\een}{\end{equation*}}

\newtheorem{remark}{Remark}

\newcommand{\cv}{\vec{c}}

\newcommand{\nv}{\vec{n}}

\newcommand{\wv}{\vec{w}}
\newcommand{\xv}{\vec{x}}
\newcommand{\yv}{\vec{y}}

\newcommand{\Dm}{\vec{D}}

\newcommand{\Gm}{\vec{G}}
\newcommand{\Hm}{\vec{H}}
\newcommand{\Id}{\vec{I}}

\newcommand{\Om}{\vec{O}}

\newcommand{\Rm}{\vec{R}}

\newcommand{\Um}{\vec{U}}

\newcommand{\Wm}{\vec{W}}
\newcommand{\Xm}{\vec{X}}

\newcommand{\Ac}{{\cal A}}

\newcommand{\Gc}{{\cal G}}

\newcommand{\Ic}{{\cal I}}

\newcommand{\Nc}{{\mathcal{N}}}
\newcommand{\Oc}{{\cal O}}

\newcommand{\Qc}{{\cal Q}}

\newcommand{\Sc}{{\cal S}}

\newcommand{\Ebb}{\mathbb{E}}
\newcommand{\Cbb}{\mathbb{C}}

\newcommand{\Rbb}{\mathbb{R}}

\newcommand{\Ibb}{\mathbb{I}}

\newcommand{\LB}{\left(}
\newcommand{\RB}{\right)}
\newcommand{\LP}{\left\{}
\newcommand{\RP}{\right\}}
\newcommand{\LSB}{\left[}
\newcommand{\RSB}{\right]}

\renewcommand{\ln}[1]{\mathop{\mathrm{ln}}\LB #1\RB}
\renewcommand{\log}[1]{\mathop{\mathrm{log}_2}\LB #1\RB}
\newcommand{\logt}[1]{\mathop{\mathrm{log}_{10}}\LB #1\RB}



%% file: tex/abstract.tex
Next-generation wireless networks aim to deliver data speeds much faster than 5G. This requires base stations with lots of antennas and a large operating bandwidth. These advanced base stations are expected to serve several multi-antenna \glspl{UE} simultaneously on the same time-frequency resources on both the uplink and the downlink. The \gls{UE} data rates are affected by the following three main factors: {\it UE rank}, which refers to the number of data layers used by each \gls{UE}, {\it \gls{UE} frequency allocation}, which refers to the assignment of slices of the overall frequency band to use for each UE in an \gls{OFDM} system, and {\it \gls{UE} power allocation/control}, which refers to the allocation of power by the base station for data transmission to each \gls{UE} on the downlink or the power used by each \gls{UE} to send data on the uplink. Since multiple \glspl{UE} are to be simultaneously served, the type of precoder used for downlink transmission and the type of receiver used for uplink reception predominantly influence these three aforementioned factors and the resulting overall \gls{UE} throughput. This paper addresses the problem of jointly selecting these three parameters specifically when \gls{ZF} precoders are used for downlink transmission  and \gls{LMMSE} receivers are employed for uplink reception. 


%% file: tex/introduction.tex
\section{Introduction}
\label{sec:intro}

Next generation wireless systems, often referred to as 6G, are being actively researched to address the ever-growing demand for data and connectivity. These systems aim to go beyond the capabilities of \gls{5GNR}, offering significantly faster speeds, ultra-low latency, and the ability to connect a vast number of devices. 6G has the potential to revolutionize various fields, from enabling real-time remote surgery to supporting entirely new applications like Extended Reality (XR).  It is expected that 6G \glspl{BS} will be equipped with a much larger number of transceiver chains in the range $128-512$ compared to $32-64$ in a typical 5G \gls{BS}, and will operate over larger bandwidths in the range $200-\SI{400}{MHz}$ compared to \SI{100}{MHz} in 5G. Each \gls{BS} is expected to serve around $10-20$ \glspl{UE} in the same time-frequency resource during peak traffic hours, and this is called {\it extreme \gls{MU-MIMO}} transmission.

To maximize the resource utilization efficiency and the \gls{UE} data rates, the following three procedures are essential: {\it \gls{UE} rank selection}, which refers to deciding on the number of data layers that a \gls{UE} uses for data transmission,  {\it \gls{UE} resource allocation}, where the system identifies which slices of the overall frequency band get assigned to a particular \gls{UE} in an \gls{OFDM} system, and {\it \gls{UE} power allocation/control}, which determines how much power a \gls{UE} uses to send data (uplink) and how much power the \gls{BS} allocates to send data to the user (downlink).

In \gls{SU-MIMO} transmissions, the \gls{OFDMA} scheme avoids inter-user interference within the same cell. However, when multiple \glspl{UE} are co-scheduled for transmission in the same time-frequency resources in a time slot, their respective transmitted data signals from the \gls{BS} (downlink) or their respective received data signals at the \gls{BS} (uplink) could potentially interfere with one another. The exact degree of this interference depends on the spatial correlation between the channels of the paired \glspl{UE}, the type of \gls{MU-MIMO} precoder used at the \gls{BS} (downlink), and the type of \gls{MU-MIMO} detector used at the \gls{BS} (uplink). Therefore, the resulting \gls{UE} data rates depend on how the frequency resources are shared between the \glspl{UE} and this in turn influences the respective \gls{UE} ranks and power allocation. Much of the literature has focused on optimizing these three procedures separately, but this leads to numerous balloon-effects where optimizing one resource reduces the gains achievable by optimizing another. Furthermore, up until the recent advent of \gls{5GNR}, \gls{MU-MIMO} was not commonly used which has also delayed the development of joint solutions for this problem. Practical solutions for the joint allocation of time-frequency resources, power control, and rank selection remain elusive and are much needed for 6G \gls{MU-MIMO} to succeed. 

This paper addresses the problem of joint resource-power allocation and rank selection for a class of linear transceivers. In particular, we consider \gls{ZF} precoders~\cite{stankovic08} for \gls{DL} transmission and \gls{LMMSE} receivers for \gls{UL} reception. We show that both these problems can be formulated under a common framework. Since the problems involve mixed integer programming for which finding the optimal solution could be computationally intensive, we provide a practically-feasible approach to obtain a reasonably good (even if possibly sub-optimal) solution. To the best of our knowledge, this is the first such work on jointly selecting the three aforementioned features based on the type of precoder or receiver used.

{\bf Related Literature}: In \gls{5GNR} and previous generations, \gls{UE} rank selection, transmit power allocation for the downlink and \gls{UE} power control for the uplink, and frequency allocation are done independently from one another and in sequence. This means that each feature operates within the constraints imposed by the decision made by the previous step. For \gls{UE} rank selection, the \gls{BS} might use the \gls{RI} sent by the \gls{UE} or select the rank through capacity optimization~\cite{chou_adaptive_rank} for both the uplink and the downlink. This is effective for \gls{SU-MIMO} but not for \gls{MU-MIMO}. 

For \gls{UE} uplink power control, the goal is to obtain good \gls{UE} rates while minimizing the effect of \gls{ICI} (which occurs due to the uplink transmissions from the \glspl{UE} in the neighbouring cells) and maintaining spectral flatness across multi-\gls{UE} receptions. The \gls{OLPC} parameters $P_0$ and $\alpha$ for a UE are configured either statically or dynamically~\cite{maggi21}. The values of these parameters ultimately specify the UL transmit power for that UE. These are effective for \gls{SU-MIMO} transmission but are not tailored to MU-MIMO transmission due to unaccounted intra-cell interference, i.e., the interference from the co-scheduled \glspl{UE} in the same cell. For frequency allocation, most commercial schedulers use round-robin frequency-domain scheduling~\cite{Xuejun_adaptive_resource_allocation} but stop short of a joint scheduling and power control solution.

For the downlink, there is a rich literature on how to design precoders (along with power allocation) for \gls{MU-MIMO}~\cite{zhao23}, but these assume that the \gls{UE} ranks and their allocated \glspl{PRB} have been fixed. \Gls{ZF} precoders~\cite{stankovic08} are more commonly used than \gls{MMSE} precoders because they suppress inter-user interference and their design does not require multiple iterations. For downlink frequency allocation,~\cite{Faroq_leasch},~\cite{NokiaDeepScheduler2024} discuss a few reinforcement learning-based techniques. However,~\cite{Faroq_leasch} is limited to  \gls{SU-MIMO} transmission with the allocation of frequency resources being independent of \gls{UE} rank and power allocation while~\cite{NokiaDeepScheduler2024} addresses joint \gls{UE} rank and resource allocation for \gls{MU-MIMO} transmission without power allocation.

{\bf Paper Organization}: The system model and a few relevant definitions are presented in Section~\ref{sec:system_model}. Section~\ref{sec:joint_res_pow_alloc} introduces the joint resource-power allocation and \gls{UE} rank selection problem while Section~\ref{sec:proposed_technique} details the proposed method that provides a practical way to obtain a reasonably good solution to this problem. Simulation results showing the efficacy of the proposed technique compared to some baseline schemes are presented in Section~\ref{sec:sim_results}, and concluding remarks constitute Section~\ref{sec:conc_remarks}.

{\bf Notation}: Boldface upper-case (lower-case) letters denote matrices (vectors). The field of complex numbers and the field of real numbers are respectively denoted by $\Cbb$ and $\Rbb$. For any set $\Sc$, $\vert \Sc \vert$ denotes its cardinality if it is finite and the notation $\Xm \in \Sc^{m \times n}$ denotes that $\Xm$ is a matrix of size $m \times n$ with each entry taking values from $\Sc$. For a matrix $\Xm$, its transpose and Hermitian transpose are respectively denoted by $\Xm^T$ and $\Xm^H$, Frobenius norm by $\Vert \Xm \Vert$, and its $(i,j)^{th}$ entry by $[\Xm]_{i,j}$. The block-diagonal matrix with diagonal blocks $\Dm_1,\Dm_2,\cdots,\Dm_n$ is denoted by $\mathrm{diag}\LSB \Dm_1,\Dm_2,\cdots,\Dm_n \RSB$, and the same notation is used for denoting a diagonal matrix with diagonal elements $d_1,\cdots,d_n$. The identity matrix and the null matrix are respectively denoted by $\Id$ and $\Om$ with their sizes understood from context. Finally, $\Ibb_{\{ A\}}$ denotes the indicator function for event $A$; $\Ibb_{\{ A\}} = 1$ if $A$ occurs, and is $0$ otherwise. 

%% file: tex/system_model.tex
\section{System Model}
\label{sec:system_model}

We consider an \gls{OFDM}-based \gls{MIMO} communication system with $n_F$ subcarriers. A \gls{RE} corresponds to a subcarrier-OFDM symbol pair in the \gls{OFDM} grid. A {\it slot} consists of $T$ \gls{OFDM} symbols (usually $T = 12$ or $14$) and a \gls{PRB} consists of 12 subcarriers.  Suppose that the \gls{BS} is equipped with $n_{B}$ antennas and serves a set of $N_{UE} \geq 1$ co-scheduled \glspl{UE}, with \gls{UE} $i$ equipped with $n_{U}^{(i)}$ antennas, $i = 1,\cdots, N_{UE}$. Each UE $i$ is assigned a rank (also called the number of data layers or spatial streams assigned for that \gls{UE}) $n_l^{(i)} \leq n_{U}^{(i)}$ with $n_L \triangleq \sum_{i=1}^{N_{UE}} n_l^{(i)} \leq n_B $. Let $\Hm_{i,f,t}\in \Cbb^{n_{B} \times n_{U}^{(i)}}$ denote the channel matrix between UE $i$ and the \gls{BS} for \gls{RE} $(f,t)$, where $f$ corresponds to the subcarrier index and $t=1,\cdots,T$, to the OFDM symbol index. While it is assumed that the \gls{BS} has perfect knowledge of this \gls{CSI}, in practice, only imperfect estimates are available, either through \gls{SRS}~\cite[Sec. 6.4.1.4]{3GPP_MCS_table_2020} transmission on the uplink in a \gls{TDD} system or through \gls{UE} \gls{CSI} feedback in a \gls{FDD} system. Moreover, these estimates are only available at a certain granularity (e.g., one channel estimate for every two consecutive \glspl{PRB}). In this paper, we call this set of contiguous \glspl{PRB} for which a single channel estimate is available a \gls{RBG}. This can also be a group of \glspl{PRB} over which the channel is approximately constant. Therefore, {\it resource allocation refers to the assignment of \glspl{RBG} to the $N_{UE}$ \glspl{UE} that are to be served in the slot in context in order to optimize a certain performance metric, with possibly multiple \glspl{UE} sharing the same \gls{RBG}}. It is assumed that once a \gls{RBG} is assigned to a \gls{UE}, all the $T$ \gls{OFDM} symbols are used for communication by the \gls{UE} in that \gls{RBG}. The constellation used for communication between the \gls{BS} and \gls{UE} $i$ is a unit energy QAM constellation of size $2^{m_i}$ for some $m_i=2,4,6,8, \cdots$, and is denoted by $\Qc_i$. 

\subsection{Uplink transmission} 
\label{subsec:UL}
Let $\Wm_{i,f,t} \in \Cbb^{n_U^{(i)} \times n_l^{(i)}}$ be the precoding matrix used by UE $i$, with each column of $\Wm_{i,f,t}$ having unit norm. Let $\Ic_{f,t} \subseteq \{1,\cdots,N_{UE}\}$ denote the set of indices of the \glspl{UE} that have been allocated the \gls{RBG} that $(f,t)$ belongs to. Suppose that UE $i \in \Ic_{f,t}$ transmits $\xv_{i,f,t} \triangleq \LSB x_{1,i,f,t},\cdots, x_{n_l^{(i)},i,f,t} \RSB^T \in \Qc_i^{n_l^{(i)} \times 1}$ on \gls{RE} $(f,t)$, with each entry of  $\xv_{i,f,t} $ taking values from $\Qc_i$. Further, let $p_{i,j,f,t} > 0$ denote the transmit power used by \gls{UE} $i$ for the $j^{th}$ layer on \gls{RE} $(f,t)$ with $p_{i,j,f,t} = 0, \forall j$, if $i \notin \Ic_{f,t}$, and $\Dm_{i,f,t} \triangleq \mathrm{diag}\LSB \sqrt{p_{i,1,f,t}},\cdots, \sqrt{p_{i,n_l^{(i)},f,t}} \RSB \in \Rbb^{n_l^{(i)}\times n_l^{(i)}}$,  . With a maximum \gls{UE} transmit power of $P_{U,max}$, it follows that 
\begin{equation}\label{eq:ul_PC}
     \sum_{f=1}^{n_F}\sum_{j=1}^{n_l^{(i)}}p_{i,j,f,t} \leq P_{U,max}, \forall i, \forall t.
\end{equation}
With $n_{L,f,t} \triangleq  \sum_{i\in \Ic_{f,t}} n_l^{(i)}$ being the total number of layers on \gls{RE} $(f,t)$, the signal model for \gls{RE} $(f,t)$ is 
\begin{equation}\label{eq:ul_sig_model}
    \yv_{f,t} = \sum_{i\in \Ic_{f,t}}\Hm_{i,f,t}\Wm_{i,f,t}\Dm_{i,f,t}\xv_{i,f,t} + \nv_{f,t} = \bar{\Hm}_{f,t}\xv_{f,t} + \nv_{f,t} 
\end{equation}
where $ \yv_{f,t} \in \Cbb^{n_B \times 1}$, $\bar{\Hm}_{f,t} \in \Cbb^{n_B \times n_{L,f,t}}$ is the effective channel matrix obtained by stacking the matrices $\{ \Hm_{i,f,t}\Wm_{i,f,t}\Dm_{i,f,t}, i \in \Ic_{f,t} \}$ column-wise, $\xv_{f,t}$ is a vector of size $n_{L,f,t}$ obtained by stacking the elements of $\{ \xv_{i,f,t}, i \in \Ic_{f,t} \}$ one below the other, and $\nv_{f,t} \in \Cbb^{n_B \times 1}$ is interference-plus-noise (including \gls{ICI}) with mean $\mathbf{0}$ and covariance matrix $\Rm_{f,t} \in \Cbb^{n_B \times n_B}$. After noise-whitening, we have  
\begin{equation}\label{eq:ul_NW_model}
    \yv_{f,t}^{\prime} \triangleq \Rm_{f,t}^{-1/2}\yv_{f,t} = \Hm_{f,t}^{\prime}\xv_{f,t} + \nv_{f,t}^{\prime} 
\end{equation}
where $ \Hm_{f,t}^{\prime} \triangleq  \Rm_{f,t}^{-1/2}\bar{\Hm}_{f,t} \in \Cbb^{n_B \times n_{L,f,t}}$, and $\nv_{f,t}^{\prime}$ has covariance $\Id$ and is approximated to be standard complex \gls{AWGN}. \Gls{LMMSE} detection~\cite[Ch. 8]{Tse2005} involves equalization using the matrix $\Gm_{f,t} \triangleq \LSB \LB\Hm_{f,t}^{\prime}\RB^H\Hm_{f,t}^{\prime } + \Id\RSB^{-1}\LB\Hm_{f,t}^{\prime}\RB^H \in \Cbb^{n_{L,f,t} \times n_B} $, resulting in 
\begin{equation}\label{eq:ZF_detection}
    \Gm_{f,t}\yv_{f,t}^{\prime} = \Gm_{f,t} \Hm_{f,t}^{\prime }  \xv_{f,t} + \Gm_{f,t}\nv_{f,t}^{\prime}. 
\end{equation}
Except for the very low \gls{SNR} regime, the performance of an \gls{LMMSE} detector is close to that of a \gls{ZF} detector~\cite[Ch. 8]{Tse2005} which uses $\Gm_{f,t} \triangleq \LSB \LB\Hm_{f,t}^{\prime}\RB^H\Hm_{f,t}^{\prime } \RSB^{-1}\LB\Hm_{f,t}^{\prime}\RB^H \in \Cbb^{n_{L,f,t} \times n_B} $. In this case, the post-equalization \gls{SINR} for the $j^{th}$ received symbol of \gls{UE} $i$, $i\in \Ic_{f,t}$, $j=1,\cdots, n_l^{(i)}$, on \gls{RE} $(f,t)$, is given as 
\begin{equation}\label{eq:post_eq_sinr_ul}
    \rho_{i,j,f,t} = \frac{1}{\LSB \LSB \LB\Hm_{f,t}^{\prime}\RB^H\Hm_{f,t}^{\prime}\RSB^{-1} \RSB_{l,l}}
\end{equation} 
where $l$ corresponds to the position of the $j^{th}$ symbol in the vector $\xv_{f,t}$. So, while the \gls{LMMSE} detector is superior to a \gls{ZF} detector, especially for ill-conditioned \gls{MIMO} channels, \eqref{eq:post_eq_sinr_ul} is still a good approximation and a simpler closed-form expression for the post-equalization \gls{SINR} obtained using an \gls{LMMSE} detector.

\subsection{Downlink transmission}
\label{subsec:DL}
 With $\Ic_{f,t}$ and $n_{L,f,t}$ as defined in the previous subsection, let $\Wm_{i, f,t} \in \Cbb^{n_B \times n_l^{(i)}}$ denote the part of the precoding matrix associated with \gls{UE} $i$ if $i\in \Ic_{f,t}$. Then, the overall precoding matrix  $\Wm_{f,t} \in \Cbb^{n_B \times n_{L,f,t}}$ used by the \gls{BS} is obtained by stacking the matrices $\{\Wm_{i, f,t}, i \in \Ic_{f,t} \}$ column-wise, with each column of $\Wm_{f,t}$ having unit norm. Further, let $a_{k,i,j,f,t} \triangleq \vert \LSB \Wm_{i, f,t} \RSB_{k,j} \vert^2$, $k=1\cdots, n_B, j = 1,\cdots, n_l^{(i)}$. The \gls{BS} transmits $\xv_{i,f,t} \triangleq \LSB x_{i,1,f,t},\cdots, x_{i,n_l^{(i)},f,t} \RSB^T \in \Qc_i^{n_l^{(i)} \times 1}$ on \gls{RE} $(f,t)$ to UE $i$ if $i \in \Ic_{f,t}$, and uses a transmit power $p_{i,j,f,t} > 0$ for the $j^{th}$ layer of \gls{UE} $i$ ($p_{i,j,f,t} = 0$ if $i \notin \Ic_{f,t}$). Let $\Dm_{i,f,t} \triangleq \mathrm{diag}\LSB \sqrt{p_{i,1,f,t}},\cdots, \sqrt{p_{i,n_l^{(i)},f,t}} \RSB$. With a maximum \gls{BS} transmit power of $P_{B,max}$ and a \gls{PAPC} of $P_{ant} = P_{B,max}/n_B$, it follows that $\forall t = 1,\cdots, T$,
\begin{eqnarray}\label{eq:dl_total_PC}
     \sum_{k=1}^{n_B}\sum_{i\in \Ic_{f,t}}\sum_{j=1}^{n_l^{(i)}}\sum_{f=1}^{n_F}a_{k,i,j,f,t}p_{i,j,f,t} & \leq & P_{B,max}, \\  \label{eq:dl_PAPC}
   \sum_{i\in \Ic_{f,t}}\sum_{j=1}^{n_l^{(i)}}\sum_{f=1}^{n_F}a_{k,i,j,f,t}p_{i,j,f,t} & \leq & P_{ant},  
\end{eqnarray}
$\forall k=1,\cdots, n_B$ in \eqref{eq:dl_PAPC}. In this paper, we consider the class of precoders called \gls{ZF} precoders~\cite{stankovic08} that satisfy, for any $j,i \in \Ic_{f,t} $, $\Hm_{j,f,t}^T\Wm_{i, f,t} = \Om$ when $j \neq i$, and $\Hm_{i,f,t}^T\Wm_{i, f,t} = \Um_{i,f,t}\mathbf{\Lambda}_{i,f,t}$ for some matrix $\Um_{i,f,t} \in \Cbb^{n_U^{(i)} \times n_l^{(i)}}$ satisfying $\Um_{i,f,t}^H\Um_{i,f,t} = \Id$ and $\mathbf{\Lambda}_{i,f,t} = \mathrm{diag}\LSB \sqrt{\lambda_{i,1,f,t}}, \cdots, \sqrt{\lambda_{i,n_l^{(i)},f,t}} \RSB$ with positive real-valued diagonal elements. With this, the signal model for the received signal at \gls{UE} $i$ on \gls{RE} $(f,t)$ is 
\begin{equation}\label{eq:dl_sig_model}
    \yv_{i, f,t} = \Hm_{i,f,t}^T\Wm_{i,f,t}\Dm_{i,f,t}\xv_{i,f,t} + \nv_{i,f,t} 
\end{equation}
where $\nv_{i,f,t} \in \Cbb^{n_U^{(i)} \times 1}$ is the interference-plus-noise with mean $\mathbf{0}$ and covariance matrix $\Rm_{i,f,t} \in \Cbb^{n_U^{(i)} \times n_U^{(i)}}$. Unlike the \gls{BS} receiver which has several receive antennas resulting in colored interference noise on the uplink, the \gls{UE} has only a few antennas, typically ranging from $2$ to $8$. Hence, $\Rm_{i,f,t} \approx \sigma_i^2 \Id$ for some positive real-valued $\sigma_i$. After receiver equalization, we have  
\begin{equation}\label{eq:dl_NW_model}
    \yv_{i,f,t}^{\prime} \triangleq \Um_{i,f,t}^H\yv_{i,f,t} = \mathbf{\Lambda}_{i,f,t}\Dm_{i,f,t}\xv_{i,f,t} + \nv_{i,f,t}^{\prime} 
\end{equation}
where $\nv_{f,t}^{\prime} \in  \Cbb^{n_l^{(i)} \times 1}$ has covariance $\sigma_i^2\Id$ and is approximated to be standard complex \gls{AWGN}. The post-equalization \gls{SINR} $ \rho_{i,j,f,t}$ for the $j^{th}$ received symbol of \gls{UE} $i$, $i \in \Ic_{f,t}$ on \gls{RE} $(f,t)$ is 
\begin{equation}\label{eq:post_eq_sinr_dl}
     \rho_{i,j,f,t} = {\lambda_{i,j,f,t}p_{i,j,f,t}}/{\sigma_i^2}.
\end{equation}

\subsection{Expected UE Rates} \label{subsec:ue_rates}
Modern communication systems use \gls{BICM}~\cite{Caire1998} where, due to the usage of scrambling and interleaving, the transmitted bits in an \gls{RE} are approximately independent and identically distributed. Let $\Gc_i \triangleq \{ (f,t) \vert i \in \Ic_{f,t}\}$ denote the indices of the \gls{RE} allocated for communication between the \gls{BS} and \gls{UE} $i$. In practical communication systems, there is finite number of \glspl{MCS}. For example, \gls{5GNR} Table 2 supports up to $27$ \gls{MCS} levels~\cite[Table 5.1.3.1-2]{3GPP_MCS_table_2020}, with the lowest level using $4$-QAM and a rate-$\frac{120}{1024}$ \gls{LDPC} code with \gls{SE} given by $\LB 120/1024\RB\log{4} = 0.234$, and the highest level using $256$-QAM and a rate-$\frac{948}{1024}$ \gls{LDPC} code with \gls{SE} $\LB948/1024\RB \log{256} = 7.40$. We denote the minimum \gls{MCS} \gls{SE} by $r_{min}$ and the maximum \gls{SE} possible by $r_{max} = \log{\vert Q_{max}\vert}$, respectively, where $ Q_{max}$ is the highest-order constellation used in the system. The achievable rate $R_{UE,i}$ (in bits per slot) for UE $i$ is upper-bounded as
\begin{equation}\label{eq:ue_rate}
    R_{UE,i} \leq  \sum_{(f,t) \in \Gc_i } \sum_{j=1}^{n_l^{(i)}}I_{Q_i}\LB \rho_{i,j,f,t} \RB 
\end{equation}
where $\rho_{i,j,f,t,i}$ is the post-equalization \gls{SINR} for the $j^{th}$ received symbol of \gls{UE} $i$ in the \gls{RE} indexed by $(f,t)$, and $I_Q(x)$ refers to the \gls{MI} between the input and the output in an \gls{AWGN} channel for constellation $Q$ and an \gls{SNR} of $x$. Since there is a minimum \gls{MCS} \gls{SE} $r_{min}$, reliable communication is not guaranteed if the expression in \eqref{eq:ue_rate} is lower than $\left\vert \Gc_i \right\vert n_l^{(i)}r_{min}$. Therefore, the achievable rate for \gls{UE} $i$ is
\begin{equation}\label{eq:ue_rate_practical}
 R_{UE,i}  =  x\Ibb_{\LP x > n_l^{(i)}\vert \Gc_i\vert r_{min} \RP}
\end{equation}
where $x$ where is given by the R.H.S. of \eqref{eq:ue_rate}.

%% file: tex/joint_resource_allocation.tex
\section{The Joint Resource-power Allocation and UE Rank selection Problem}
\label{sec:joint_res_pow_alloc}

\begin{figure}[!t]
      \centering
      \includegraphics[width=3.2in]{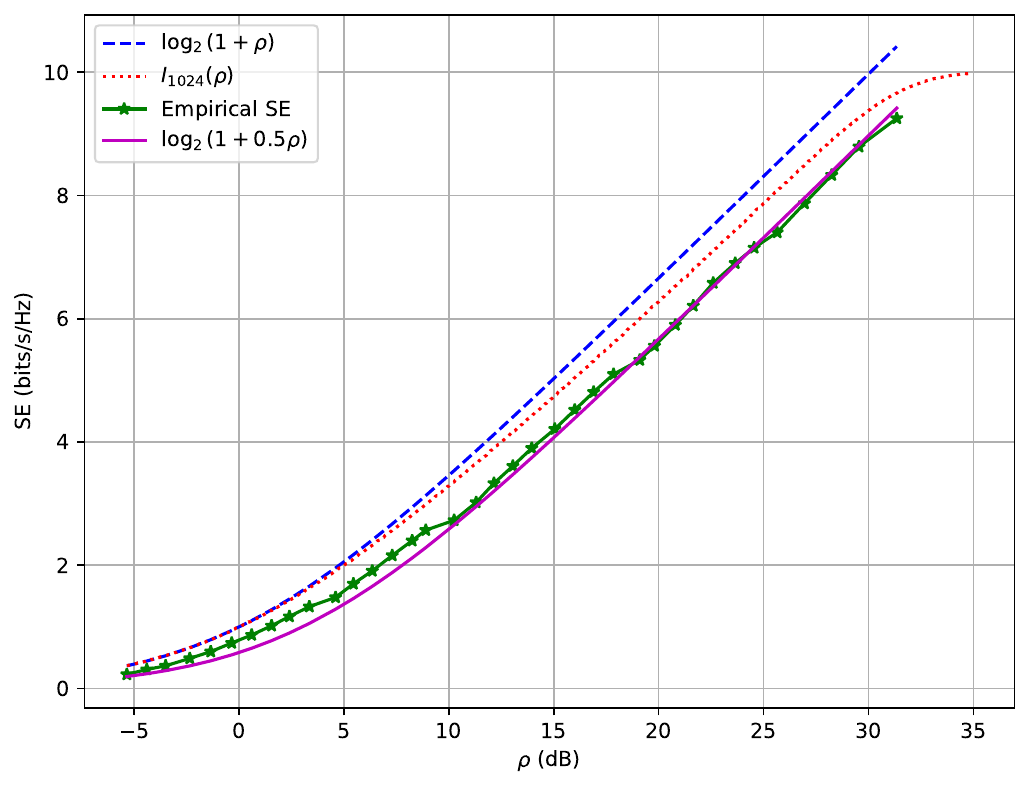}
      \caption{Comparison between the plots of \gls{SE} vs. \gls{SNR} for different functions.}
      \label{fig:log_approx}
\end{figure}

For casting the joint resource-power allocation and \gls{UE} rank selection as an optimization problem, we first need to obtain a suitable differentiable function that well-approximates $I_{Q_{max}}(.)$ in \eqref{eq:ue_rate}, where $Q_{max}$ is the highest-order constellation used. Fig.~\ref{fig:log_approx} plots the \gls{SE} as a function of \gls{SNR}, denoted by $\rho$, for four functions; $\log{1+\rho}$ is the Shannon capacity, $I_{1024}(\rho)$ refers to the empirically-obtained mutual information between the input and output of an \gls{AWGN} channel with $1024$-QAM, ``Empirical SE" refers to the empirically obtained set of values of \glspl{SNR} required to achieve a \gls{CER} of $10^{-3}$ in a \gls{SISO} \gls{AWGN} channel for each \gls{MCS} level of the combined tables of~\cite[Tables 5.1.3.1-2,  5.1.3.1-4]{3GPP_MCS_table_2020} using \gls{5GNR} \gls{LDPC} codes of length $2880$. As evident from the figure, the Shannon capacity is an overly optimistic approximation for practical communication systems that use finite-sized constellations (e.g., $2^m$-QAM) while $I_{1024}(\rho)$ is still too optimistic for systems that use practical error-correcting codes. A good approximation needs to be conservative for lower \gls{MCS} indices due to the fact that lower \gls{MCS} levels serve \glspl{UE} that have relatively low \gls{SINR}, and $\log{1+0.5\rho}$ is one such logarithmic approximation that is closer to practical rates than the theoretical mutual information. We would like to emphasize that $\log{1+0.5\rho}$ is just an exemplary function that will be used in the rest of this paper, and better differentiable functions can be obtained by suitable optimization.  

As mentioned in the previous section, the \gls{BS} needs to have reasonably good channel estimates to perform resource-power allocation. These channel estimates are not available for each \gls{PRB}, but for groups of \glspl{PRB} which we call \gls{RBG} in this paper. Let $n_{PRB}$ denote the total number of \glspl{PRB} and $n_{RBG}$ the total number of \glspl{RBG} in the system so that we have a channel estimate for every $n_{PRB}/n_{RBG}$ \glspl{PRB}. Let $\delta_{i,g} \in \{ 0, 1\}$ indicate whether \gls{RBG} $g$ is allocated to \gls{UE} $i$ ($\delta_{i,g} = 1$) or not ($\delta_{i,g} = 0$). Note that unlike in \gls{5GNR}, we do not restrict the frequency allocation to be contiguous in this paper. For every notation used in Section~\ref{sec:system_model}, we use the subscript $g$ instead of $(f,t)$ to denote that the notation applies to \gls{RBG} $g$ instead of \gls{RE} $(f,t)$. With this, the \gls{BS} uses the channel estimates to obtain estimates of the \gls{SINR} $\rho_{i,j,g}$ for the $j^{th}$ symbol of \gls{UE} $i$ on \gls{RBG} $i$ using \eqref{eq:post_eq_sinr_ul} for the uplink and \eqref{eq:post_eq_sinr_dl} for the downlink,  where, for the latter, the \gls{BS} uses the interference-plus-noise variance feedback from each \gls{UE}. Henceforth, we use the notation $\Gc \triangleq \LP1,\cdots,n_{RBG} \RP $, $\Nc_{UE} \triangleq \LP 1,\cdots, N_{UE}\RP$. 

Let $R_{UE,i}$ denote the normalized (with respect to some constant) rate in bits per slot for \gls{UE} $i$. It is a common practice to maximize a weighted $\alpha$-fair utility function~\cite{Uchida_alpha_fair} $\sum_{i=1}^{N_{UE}}a_i f_{\alpha}\LB R_{UE,i} \RB$, where $f_{\alpha}(.)$, $\alpha > 0$, is an increasing, strictly concave, and continuously differentiable function on the open interval $\LB0, \infty \RB$, as follows: 
\begin{equation}
 f_{\alpha}\LB x \RB = \begin{cases}
    \frac{x^{1-\alpha}}{1-\alpha}, & \textrm{ if }\alpha > 0, \alpha \neq 1, \\
    \ln{x}, & \textrm{ if } \alpha = 1. 
\end{cases} 
\end{equation}
Here, the weights satisfy $a_i > 0, \forall i$. In this paper, we take $a_i = 1, \forall i \in \Nc_{UE}$ and $\alpha = 1$, which corresponds to the {\it \gls{GM} rate optimization or proportional fairness}~\cite{Kelly1998RateCF}. However, our problem formulation is applicable to any value of $\alpha$.

\subsection{Uplink Joint Resource-power Allocation and UE Rank Selection}
\label{subsec:ul_problem}

The \gls{BS} obtains the channel estimates $\hat{\Hm}_{i,g} \in \Cbb^{n_B \times n_l^{(i)}}$ corresponding to the first $n_l^{(i)}$ columns of $\Rm_{g}^{-1/2}\Hm_{i,f,t}\Wm_{i,f,t}$ of Section~\ref{subsec:UL} for \gls{RBG} $g$ (note that the interference-plus-noise covariance is now calculated at the \gls{RBG} level and not at the \gls{RE} level). Let $\hat{\Hm}_{g} \triangleq \LSB \hat{\Hm}_{1,g},\cdots, \hat{\Hm}_{N_{UE},g} \RSB \in \Cbb^{n_B \times n_L}$ where $n_L = \sum_{i=1}^{N_{UE}} n_l^{(i)}$. Then, with transmit powers $p_{i,j,g} > 0$, $j=1,\cdots,n_l^{(i)}$, for \gls{UE} $i$ on \gls{RBG} $g$, we get an \gls{SINR} of $\rho_{i,j,g} = \lambda_{i,j,g}p_{i,j,g}$ for a hypothetically transmitted $j^{th}$ layer of UE $i$ on \gls{RBG} $g$, where $\lambda_{i,j,g}$ is the reciprocal of the $(\sum_{i^{\prime} = 1}^{i-1}n_l^{(i^{\prime})} + j)^{th}$ diagonal element of $\LSB  \hat{\Hm}_{g}^H\hat{\Hm}_{g}\RSB^{-1} $. Therefore, from the log approximation function $\log{1+0.5\rho}$, the normalized expected rate in bits per slot (normalized by $T$ times the number of subcarriers in a \gls{RBG}) for \gls{UE} $i$ is given by
\begin{equation}\label{eq:ul_UE_rate_log_approx}
R_{UE,i} = \sum_{g=1}^{n_{RBG}}\sum_{j=1}^{n_l^{(i)}}\log{1+0.5\delta_{i,g}\lambda_{i,j,g}p_{i,j,g}}.
\end{equation}
Let $\rho_{max}$ correspond to the \gls{SNR} required to achieve a SE of $r_{max}$, i.e., $\rho_{max} = 2\LB2^{r_{max}}-1\RB$. With this, the goal of the resource allocation problem is to maximize the \gls{GM} of the \gls{UE} rates, or equivalently,  $\sum_{i=1}^{N_{UE}} \ln{R_{UE,i}}$, as follows:
\begin{subequations}
\begin{alignat}{3} \nonumber
        & P_{UL}:\\ \label{eq:ul_opt_problem}
        &\underset{p_{i,j,g}, \delta_{i,g}, n_l^{(i)}}{\max} &&  \sum_{i=1}^{N_{UE}} \ln{R_{UE,i}} \\  \label{eq:ul_opt_problem_cons_1}
        & \text{s.t.} & & \delta_{i,g} \in \{0,1\}, \sum_{g \in \Gc} \delta_{i,g} \geq 1,\forall i, g,\\ \label{eq:ul_opt_problem_cons_1a}
        &  & &  n_l^{(i)} \leq   n_U^{(i)}, \forall i,\\ 
         \label{eq:ul_opt_problem_cons_2}
        & & & 0 \leq p_{i,j,g} \leq {\rho_{max}}/{\lambda_{{i,j,g}}}, \forall i, \forall g, \forall j, \\
        \label{eq:ul_opt_problem_cons_3}
         & & & R_{UE,i} \geq r_{min} n_l^{(i)}\sum_{g \in \Gc}\delta_{i,g}, \forall i,  \\ 
        \label{eq:ul_opt_problem_cons_4}
        & & & \sum_{g \in \Gc}\sum_{j=1}^{n_l^{(i)}}\delta_{i,g}p_{i,j,g} \leq P_{U,max}, \forall i.
\end{alignat}
\end{subequations}

In the above optimization problem, \eqref{eq:ul_opt_problem_cons_1} imposes an allocation requirement of at least one \gls{RBG} for each \gls{UE} but any arbitrary value $n_{RBG,min} \geq 1$ is possible. Also, \eqref{eq:ul_opt_problem_cons_2} can be replaced by a \gls{UE}-specific maximum power that would cater to power allocation for \glspl{UE} with different \glspl{QoS}, or to limit the \gls{ICI}. {\bf In the latter case, $\rho_{max}$ is the equivalent of $(P_0, \alpha)$ for uplink power control} (see Section \ref{subsec:uplink_sim}). The constraint in \eqref{eq:ul_opt_problem_cons_3} is a reformulation of \eqref{eq:ue_rate_practical}. 

\subsection{Downlink Joint Resource-power Allocation and UE Rank Selection}
\label{subsec:dl_problem}

The setup is as described in Section~\ref{subsec:DL} with the usage of notation and the \gls{RBG} indices in the subscripts of symbols as explained in Section~\ref{subsec:ul_problem}. Without loss of generality, we assume that $\sigma_i^2 = 1$, $\forall i=1,\cdots, N_{UE}$, so that from \eqref{eq:post_eq_sinr_dl}, the normalized expected rate in bits per slot (normalized by $T$ times the number of subcarriers in a \gls{RBG}) for \gls{UE} $i$ is the same as in \eqref{eq:ul_UE_rate_log_approx}. The resource allocation problem $P_{DL}$ can now be expressed as 
\begin{subequations}
\begin{alignat}{3}\nonumber
         &P_{DL}: \\ \label{eq:dl_opt_problem}
          & \underset{p_{i,j,g}, \delta_{i,g}, n_l^{(i)}}{\max}  &\quad&  \sum_{i=1}^{N_{UE}} \ln{R_{UE,i}} \\  \label{eq:dl_opt_problem_cons_1}
         & \text{s.t.} & & \delta_{i,g} \in \{0,1\}, \sum_{g \in \Gc} \delta_{i,g} \geq 1, \forall i, \forall g,\\
         \label{eq:dl_opt_problem_cons_1a}
         & & &  n_l^{(i)} \leq   n_U^{(i)}, \forall g,\\ 
         \label{eq:dl_opt_problem_cons_2}
       & & & 0 \leq p_{i,j,g} \leq {\rho_{max}}/{\lambda_{{i,j,g}}}, \forall i, \forall g, \forall j, \\
        \label{eq:dl_opt_problem_cons_3}
        & & &  R_{UE,i} \geq r_{min} n_l^{(i)}\sum_{g \in \Gc}\delta_{i,g}, \forall i, \\ 
        \label{eq:dl_opt_problem_cons_4}
        & &&  \sum_{k=1}^{n_B}\sum_{i=1}^{N_{UE}}\sum_{j=1}^{n_l^{(i)}}\sum_{g \in \Gc}\delta_{i,g}a_{k,i,j,g}p_{i,j,f,t} \leq P_{B,max},\\
        \label{eq:dl_opt_problem_cons_5}
        & & &  \sum_{i=1}^{N_{UE}}\sum_{j=1}^{n_l^{(i)}}\sum_{g \in \Gc}\delta_{i,g}a_{k,i,j,g}p_{i,j,g}  \leq P_{ant}, \forall k.
\end{alignat}
\end{subequations}

Except for the expressions for the post-equalization \gls{SINR}, the only other significant difference between the problem formulations $P_{UL}$ and $P_{DL}$ is that for the downlink, the available \gls{BS} transmit power needs to be shared by all the \gls{UE} as shown in \eqref{eq:dl_opt_problem_cons_4}, and that there is also an additional \gls{PAPC} at the \gls{BS} as shown in \eqref{eq:dl_opt_problem_cons_5}.

Both $P_{UL}$ and $P_{DL}$ are non-convex optimization problems and involve mixed integer programming. It is possible that there exists no unique optimal solution, or that an optimal solution cannot be obtained without too many iterations. In the next section, we propose a method to obtain a reasonably good (even if possibly sub-optimal) solution to the two problems without too many iterations.

%% file: tex/proposed_technique.tex
\section{The Proposed Optimization Technique}
\label{sec:proposed_technique}

\gls{SINR} estimation as given by \eqref{eq:post_eq_sinr_ul} and \eqref{eq:post_eq_sinr_dl} requires computationally intensive matrix inversions, potentially involving Moore-Penrose inversion, \gls{SVD} or \gls{EVD} depending on the \gls{ZF}-precoding type used~\cite{stankovic08}. Since $P_{UL}$ and $P_{DL}$ are both non-convex, there is no guarantee that any proposed approach might lead to an optimal solution. Hence, the goal is to obtain a possibly sub-optimal solution without incurring a high computational complexity. In this regard, we split both optimization problems into two convex optimization stages with an intermediate processing stage in between. In lieu of $P_{UL}$ and $P_{DL}$, we consider the following problems where $\delta_{i,g}$ and $n_{l}^{(i)}$ are fixed and not optimized over:
\begin{subequations}
\begin{alignat}{3} \nonumber
         & \bar{P}_{UL}: \\  \label{eq:ul_convopt_problem} 
  & \underset{p_{i,j,g}}{\min} &\quad& -\sum_{i=1}^{N_{UE}} \ln{R_{UE,i}} \\  
         \label{eq:ul_convopt_problem_cons_1}
        & \text{s.t.} &  & 0 \leq p_{i,j,g} \leq {\rho_{max}}/{\lambda_{{i,j,g}}}, \forall i, \forall g, \forall j, \\        
        \label{eq:ul_convopt_problem_cons_2}
        & & &\sum_{g=1}^{n_{RBG}}\sum_{j=1}^{n_l^{(i)}}\delta_{i,g}p_{i,j,g} \leq P_{U,max}, \forall i \in \Nc_{UE},  
\end{alignat}
\end{subequations}
for the uplink, and
\begin{subequations}
\begin{alignat}{3} \nonumber
        & \bar{P}_{DL}: \\ \label{eq:dl_convopt_problem}
         & \underset{p_{i,j,g}}{\min}  &\quad& -\sum_{i=1}^{N_{UE}} \ln{R_{UE,i}} \\  \label{eq:dl_convopt_problem_cons_1}        
        & \text{s.t.} &  & 0 \leq p_{i,j,g} \leq {\rho_{max}}/{\lambda_{{i,j,g}}}, \forall i, \forall g , \forall j, \\
       \label{eq:dl_convopt_problem_cons_2}
        & & &\sum_{k=1}^{n_B}\sum_{i=1}^{N_{UE}}\sum_{j=1}^{n_l^{(i)}}\sum_{g=1}^{n_{RBG}}\delta_{i,g}a_{k,i,j,g}p_{i,j,f,t} \leq P_{B,max},\\
        \label{eq:dl_convopt_problem_cons_3}
        & & & \sum_{i=1}^{N_{UE}}\sum_{j=1}^{n_l^{(i)}}\sum_{g=1}^{n_{RBG}}\delta_{i,g}a_{k,i,j,g}p_{i,j,g}  \leq P_{ant}, \forall k,
\end{alignat}
\end{subequations}
for the downlink. Also to note is that the non-convex constraints \eqref{eq:ul_opt_problem_cons_3} and \eqref{eq:dl_opt_problem_cons_3} are not considered. With this, both  $\bar{P}_{UL}$ and $\bar{P}_{DL}$ are now convex-optimization problems with convex objectives and affine constraints and can be solved using interior point methods~\cite{dikin1967iterative} for which there exist efficient large-scale optimization software like \gls{IPOPT}~\cite{Wachter2006}. Let $\textrm{OPT}\LB \bar{P}\LB \LP \lambda_{i,j,g},  \delta_{i,g}, n_l^{(i)}, \forall i,j,g \RP \RB \RB$ denote the solution to the convex optimization problem $\bar{P}$, where $\bar{P}$ is either $\bar{P}_{UL}$ or $\bar{P}_{DL}$ as the case may be. Further, we denote by  $f\LB \hat{\Hm}_{g}, \LP \delta_{i,g}, n_l^{(i)}, \forall i \in \Nc_{UE} \RP \RB$ the set of values $\lambda_{i,j,g}, \forall i,j,g$. Note that for the uplink (see Sec. \ref{subsec:ul_problem}), $\lambda_{i,j,g}$ is the reciprocal of the $(\sum_{i^{\prime} = 1}^{i-1}n_l^{(i^{\prime})} + j)^{th}$ diagonal element of $\LSB  \hat{\Hm}_{g}^H\hat{\Hm}_{g}\RSB^{-1} $, while for the downlink, it is obtained from the product of the channel matrix and the \gls{ZF} precoder (see Sec. \ref{subsec:DL}). In both cases, $\lambda_{i,j,g}$ depends on the set of \glspl{UE} that share the \gls{RBG} and the number of layers for each such \gls{UE}. Therefore, we add the following definition, $\forall i \in \Nc_{UE},\forall g \in \Gc$:
\begin{eqnarray} \label{eq:cond_lambda_1}
   \lambda_{i,j,g} & = & 0, \forall j \in \LP 1,\cdots, n_U^{(i)} \left \vert \delta_{i,g} = 0 \right.\RP, \\ \label{eq:cond_lambda_2}
  \lambda_{i,j,g} & = & 0,  n_l^{(i)} < j \leq  n_U^{(i)} .
\end{eqnarray}
Since the optimization problems $P_{UL}$ and $P_{DL}$ are similarly structured, we now provide a unified approach for the joint resource-power allocation and \gls{UE} rank selection for both the uplink and the downlink. In the first stage, we propose to fix $\delta_{i,g}=1$, $n_l^{(i)} = n_U^{(i)}$, $\forall i \in \Nc_{UE}, \forall g \in \Gc$, and solve $\bar{P} = \bar{P}_{UL}$ or $\bar{P}_{DL}$ as the case may be. Let $\LP p_{i,j,g}^{(1)}, j=1,\cdots, n_U^{(i)},\forall i \in \Nc_{UE}, \forall g \in \Gc \RP = \textrm{OPT}\LB \bar{P}\LB \LP \lambda_{i,j,g},  \delta_{i,g}=1, n_l^{(i)} = n_U^{(i)}, \forall i,j,g \RP \RB \RB $ be the solution obtained in Stage 1. 

\begin{algorithm}
\caption{Pseudocode for MU-MIMO resource-power allocation and \gls{UE} rank selection.}
\label{alg:pseudocode}
\begin{algorithmic}[1]    
    \Output
        \Desc{$\delta_{i,g}^*$}{: \gls{RBG} $g$ allocation indicator for \gls{UE} $i$, $\forall i \in \Nc_{UE}$, $\forall g \in \Gc$ } 
        \Desc{$ n_l^{(i)} $}{: The assigned rank for \gls{UE} $i$, $\forall i \in \Nc_{UE}$ }  
        \Desc{$p_{i,j,g}^{*} $}{: The allocated powers, $j=1\cdots, n_U^{(i)}, \forall i \in \Nc_{UE}, \forall g \in \Gc$}   
    \EndOutput
    \State $\delta_{i,g} \gets 1, n_l^{(i)} \gets  n_U^{(i)}, \forall i \in \Nc_{UE}, \forall g \in \Gc$ 
    \State $\lambda_{i,j,g} \gets f\LB \hat{\Hm}_{g}, \LP \delta_{i,g}, n_l^{(i)} , \forall i \in \Nc_{UE} \RP \RB, \forall i \in \Nc_{UE}, \forall g \in \Gc$ 
    \State $\LP p_{i,j,g}^{(1)},\forall i,j,g \RP  \gets \textrm{OPT}\LB \bar{P}\LB \LP \lambda_{i,j,g},  \delta_{i,g}, n_l^{(i)}, \forall i,j,g \RP \RB \RB$ 
    \ForAll{$i \in \Nc_{UE}$}
        \State $r_{i,j,g} \gets \log{1+0.5\lambda_{i,j,g}p_{i,j,g}^{(1)}}$, $\forall j=1,\cdots,n_U^{(i)}, \forall g \in \Gc$ 
        \State $\delta_{i,j,g} \gets 1 \textrm{ if }r_{i,j,g} \geq r_{min}$, $0$ otherwise , $\forall j=1,\cdots,n_U^{(i)}, \forall g \in \Gc$ 
        \State $r_{current} \gets 0, r_{mean} \gets 0, n_{l}^{(i)} \gets 1$
        \State $\delta_{i,g} \gets \prod_{j=1}^{n_{l}^{(i)}} \delta_{i,j,g}$, $\forall g \in \Gc$ \label{marker1}
        \State $r_{i,sum} \gets \sum_{g=1}^{n_{RGB}}\delta_{i,g}\sum_{j=1}^{ n_{l}^{(i)}}  r_{i,j,f}$, $r_{i,mean} \gets {r_{i,sum}}/\LB{n_{l}^{(i)}\sum_{g=1}^{n_{RGB}}\delta_{i,g} }\RB$
        \If{$r_{i,sum} > r_{current} \And r_{i,mean} > r_{min} \And  n_{l}^{(i)} <  n_{U}^{(i)}$ } 
           \State $ n_{l}^{(i)} \gets  n_{l}^{(i)} + 1$
           \State \Goto{marker1}
        \Else 
           \If{$\sum_{g \in \Gc}\delta_{i,g} < n_{RBG,min} $}
           \State /* {\textit{Identify a set $\Ac_{i}$ of \glspl{RBG} with the $n_{RBG,min}$ highest values of UE rates}} */
               \State $\Ac_{i} \gets \underset{\Ac \subset \Gc, \vert \Ac \vert = n_{RBG,min}}{\arg \max} \LP \sum_{g \in \Ac}\sum_{j=1}^{n_{l}^{(i)}}r_{i,j,g} \RP $
               \State $\delta_{i,g}^* \gets 1, \forall g \in \Ac_{i}$, and $\delta_{i,g}^* \gets 0, \forall g \notin \Ac_{i}$
           \Else 
               \State $\delta_{i,g}^* \gets \delta_{i,g}$
           \EndIf         
        \EndIf
    \EndFor
    \State $\lambda_{i,j,g} \gets f\LB \hat{\Hm}_{g}, \LP \delta_{i,g}^{*}, n_l^{(i)}, \forall i \in \Nc_{UE} \RP \RB$, $\forall j = 1, \cdots, n_U^{(i)}$, $\forall i \in \Nc_{UE} $, $\forall g \in \Gc$    
    \State $\LP p_{i,j,g}^{*},\forall i,j,g \RP  \gets \textrm{OPT}\LB \bar{P}\LB \LP \lambda_{i,j,g},  \delta_{i,g}^*, n_l^{(i)}, \forall i,j,g \RP \RB \RB$ 
\end{algorithmic}
\end{algorithm}

The details of the second stage are better explained using Algorithm~\ref{alg:pseudocode}. Steps 9--10 identify the weak \glspl{RBG} and layers for each \gls{UE}. Next, the idea is to start conservatively at $n_l^{(i)} = 1$ layer for each UE and check iteratively if additional layers can be accommodated. Steps 11–16 perform these operations, noting that the number of layers remains constant across all allocated \glspl{RBG} for each \gls{UE}. Step 13 is used to ensure that \eqref{eq:ul_opt_problem_cons_3} or \eqref{eq:dl_opt_problem_cons_3} is respected. Once the number of layers has been identified for each \gls{UE}, the next check is to ensure that the minimum number of \glspl{RBG} $n_{RBG,min}$ is assigned to each \gls{UE}. If not, Steps 20--21 assign the \glspl{RBG} with the $n_{RBG,min}$ highest values of estimated \gls{UE} rates for the \gls{UE} in context. Finally, once the \gls{RBG} allocation and \gls{UE} ranks have been arrived at, the values of $\lambda_{i,j,g}$ are recalculated in Step 27 if there is a change in the \gls{RBG} allocation and the \gls{UE} ranks with respect to Stage 1, keeping also in mind \eqref{eq:cond_lambda_1} and \eqref{eq:cond_lambda_2}. Step 28 ensures that the \glspl{RBG} that are not allocated and the \gls{UE} layers that are not selected will play no further part in the optimization. If each \gls{UE} is {\it servable} (i.e., it is possible to guarantee a minimum rate for that \gls{UE} for some feasible resource-power allocation), by choosing to maximize the \gls{GM} rates and due to Steps 11-16, we would have already ensured that the finally allocated \glspl{RBG} and layers for each \gls{UE} are such that each \gls{UE} has a minimum rate guarantee in the slot. Next, with $\rho_{min} \triangleq 2\LB 2^{r_{min}}-1 \RB$, we set the lower limit on $p_{i,j,g}$ in \eqref{eq:ul_convopt_problem_cons_1} and \eqref{eq:dl_convopt_problem_cons_1} to $\rho_{min}/\lambda_{i,j,g}$ if $\delta_{i,g}^* = 1$ and $j=1,\cdots,n_{l}^{(i)}$, and $0$ otherwise. Then, the output of Step 27 provides the final allocated powers to each layer and each \gls{RBG} for each \gls{UE}. 

\begin{remark}
    If there is a non-servable \gls{UE}, (i.e., a \gls{UE} with extremely poor channel conditions for which no allocations are possible to guarantee a minimum rate in the slot in context), such a \gls{UE} would adversely affect the optimization problem. A prudent approach would be to not serve this \gls{UE} at all in the first place until its channel conditions improve.
\end{remark}
 
\subsection{Uplink Scheduling Information}
\label{subsec:ul_practical_consideration}

Having obtained the \gls{UE} transmit powers, the \gls{UE} ranks, and the resource allocation indices, the \gls{BS} must now transmit this information to each co-scheduled \gls{UE} on a \gls{DCI}~\cite{3GPP_dci_2020} message. However, the standard \gls{DCI} formats in \gls{5GNR} can only encode a single low-resolution closed-loop power control command per \gls{UE} and slot. Instead, the \gls{MU-MIMO} power-control solution we propose requires \gls{TPC} commands that are \gls{RBG}-, slot-, and layer-specific. \gls{5GNR} \gls{DCI} formats are therefore insufficient to convey these data-rich \gls{BS} decisions. For this reason, we propose the following additional features to support \gls{MU-MIMO} \gls{TPC} commands. Note that these features differ in the amount of signaling overhead, and consequently, not all of them may be preferred in all scenarios.
\begin{enumerate}
    \item {\it High granularity \gls{DCI}}: The \gls{DCI} encodes the discretized full set of transmit powers for each \gls{RBG} on each layer. This incurs the highest overhead and might be practical in 6G deployments with extreme bandwidths.
    \item {\it Medium granularity \gls{DCI}}: A condensed set of discretized transmit powers, only one per layer, is sent. In this case, the \gls{UE} uses the same transmit power for all scheduled \glspl{RBG} per layer. This embodiment may yield the best trade-off between performance and signaling overhead, and it may be optimal in deployments with low-frequency selectivity, such as rural outdoors.
    \item {\it Low granularity \gls{DCI}}: A single discretized transmit power per UE is transmitted on the \gls{DCI}. In this case, the UE uses a common transmit power for all the scheduled PRBs and layers. This incurs the least overhead, and depending on the case, might not significantly degrade the performance compared to the having separate transmit powers per \gls{RBG} and layer.
\end{enumerate}

\subsection{Possible Limitations of the proposed technique}
\label{subsec:limitation}

The performance of the proposed technique hinges entirely on the accuracy of channel estimates $\hat{\Hm}_g$, $\forall g \in \Gc$. As a result, one can expect it to break down under one or more of the following scenarios.

If the \gls{BS}-\gls{UE} communication is \gls{FDD}, the uplink channel is unlikely to be the reciprocal of the downlink channel and it would not be realistic to use the proposed technique for the downlink if the channel estimates are obtained from the uplink \gls{SRS}. However, if the \gls{UE} were to feedback the downlink channel that it estimates, for example, using \glspl{CSI-RS}~\cite[Section~7.4.1.5]{3GPP_MCS_table_2020}, the proposed technique can be still employed.  
 
If some of the \glspl{UE} have poor channel conditions (e.g., cell-edge \glspl{UE}) and an insufficient power budget to ensure a certain \gls{SINR} at the \gls{BS}, the \gls{BS} would not be able to obtain good quality channel estimates from the \glspl{SRS} of these \glspl{UE}. This scenario would adversely impact the downlink more because the assumed \gls{ZF} precoder design would no longer be valid. 

\subsection{Computational Complexity and Practical Considerations}
\label{subsec:comp_complexity}
Interior-point methods solve the following convex optimization problem in its standard form~\cite{Wachter2006}:
\begin{subequations}
\begin{alignat}{2}
    \label{eq:std_conv_opt}
    & \min_{\xv \in \Rbb^{n}} &\qquad& f(\xv) \\
    &\text{subject to} & & \cv_L \leq c(\xv) \leq  \cv_U, \\
   &  & & \xv_L \leq \xv \leq \xv_U,   
   \end{alignat}
\end{subequations}
where $\xv_L \in [ -\infty, \infty)^n$, $\xv_U \in ( -\infty, \infty ]^n$ with $\LSB \xv_L \RSB_i <  \LSB \xv_U \RSB_i$, $\forall i=1,\cdots,n$, $\cv_L \in [ -\infty, \infty)^m$, $\cv_U \in ( -\infty, \infty ]^m$ with $\LSB \cv_L \RSB_i <  \LSB \cv_U \RSB_i$, $\forall i=1,\cdots,m$. The objective function is $f : \Rbb^{n} \to \Rbb$ and the constraint function is $c : \Rbb^{n} \to \Rbb^{m}$ (with any equality constraint being set by choosing $\cv_L = \cv_U$). For a tolerance $\epsilon > 0$, the number of required iterations to obtain a solution is $\Oc\LB \sqrt{n}\ln{1/\epsilon} \RB$~\cite{wright_IP_methods}. Further, each iterations involves solving a linear system of equations that has a worst-case complexity of $\Oc \LB n^3 \RB$. Note that this is an upper bound and dependent on the form of the objective function and how entangled the variables are with one another, and most practical linear solvers have a complexity less than $\Oc \LB n^3 \RB$. Nevertheless, the worst-case complexity of any interior-point method in order to achieve a tolerance of $\epsilon > 0$ is $\Oc\LB n^{3.5}\ln{1/\epsilon} \RB$. 

The proposed technique uses two optimization stages with each optimization stage solving a convex optimization problem with $n = n_{RBG}n_U$ variables where $n_U = \sum_{i}n_{U}^{(i)}$. For the uplink problem \eqref{eq:ul_convopt_problem}, the nature of the objective and the constraints imply that the optimization for each UE can be done separately. So, this translates to solving $N_{UE}$ independent optimization problems with $n = n_{RBG}n_{U}^{(i)}$ variables for UE $i$ in the first stage and $n = n_{RBG}n_{l}^{(i)}$ variables in the second stage. But each stage of the proposed technique requires the computation of $\LSB  \hat{\Hm}_{g}^H\hat{\Hm}_{g}\RSB^{-1} $, $\forall g=1,\cdots,n_{RBG}$, where $\hat{\Hm}_{g} \in \Cbb^{n_B \times n_U}$, which has a worst-case complexity of $\Oc \LB n_{RBG} n_U^3 \RB$. So, the worst-case complexity of the proposed technique for the uplink is $\Oc \LB N_{UE} \LB n_{RBG} n_{U,max}\RB^{3.5}\ln{1/\epsilon}  +  n_{RBG} n_U^3 \RB$, where $n_{U,max} = \max_{i\in \Nc_{UE}} \{ n_{U}^{(i)} \}$.

For the downlink, all the $n = n_{RBG}n_U$ variables need to be jointly optimized, leading to a worst-case complexity of $\Oc \LB \LB n_{RBG} n_{U}\RB^{3.5}\ln{1/\epsilon} \RB$. Note that while the \gls{ZF} precoders need to be calculated in order to obtain the $\lambda_{i,j,g}$, this is anyway unavoidable for downlink transmission and hence not part of the additional complexity resulting from the usage of the proposed technique. Further, note that since the Hessian of the objective function as shown in \eqref{eq:dl_convopt_problem} allows a block-diagonal structure, each iteration of the optimization problem involves solving a sparse linear system of equations which significantly reduces the overall complexity. If \gls{RBG} allocation is not considered, i.e., all the available \glspl{RBG} are allocated to every \gls{UE}, the complexity is significantly reduced from $\Oc \LB \LB n_{RBG} n_{U}\RB^{3.5}\ln{1/\epsilon} \RB$ to $\Oc \LB n_{U}^{3.5}\ln{1/\epsilon} \RB$, but this might come at the cost of throughput performance, especially in highly frequency selective channels. Further, we can always neglect the requirement on a certain error tolerance and fix the number of iterations to a predetermined number. This would further tradeoff performance for lower complexity.

The computational burden of our method, requiring the execution of \gls{IPOPT} every slot, presents a substantial challenge for real-time 6G implementation. To address this, we propose investigating both supervised and reinforcement learning approaches. Supervised learning could involve training a model on a dataset for which the labels have been obtained by the proposed technique, allowing for rapid prediction in subsequent slots. Alternatively, reinforcement learning, guided by our proposed technique as an expert policy, could uncover even better performing methods. Future research will focus on exploring these ML-based techniques, aiming to balance computational efficiency and performance.

%% file: tex/simulation_results.tex
\section{Simulation Results}
\label{sec:sim_results}

\begin{figure}[!t]
      \centering
      \includegraphics[width=3in]{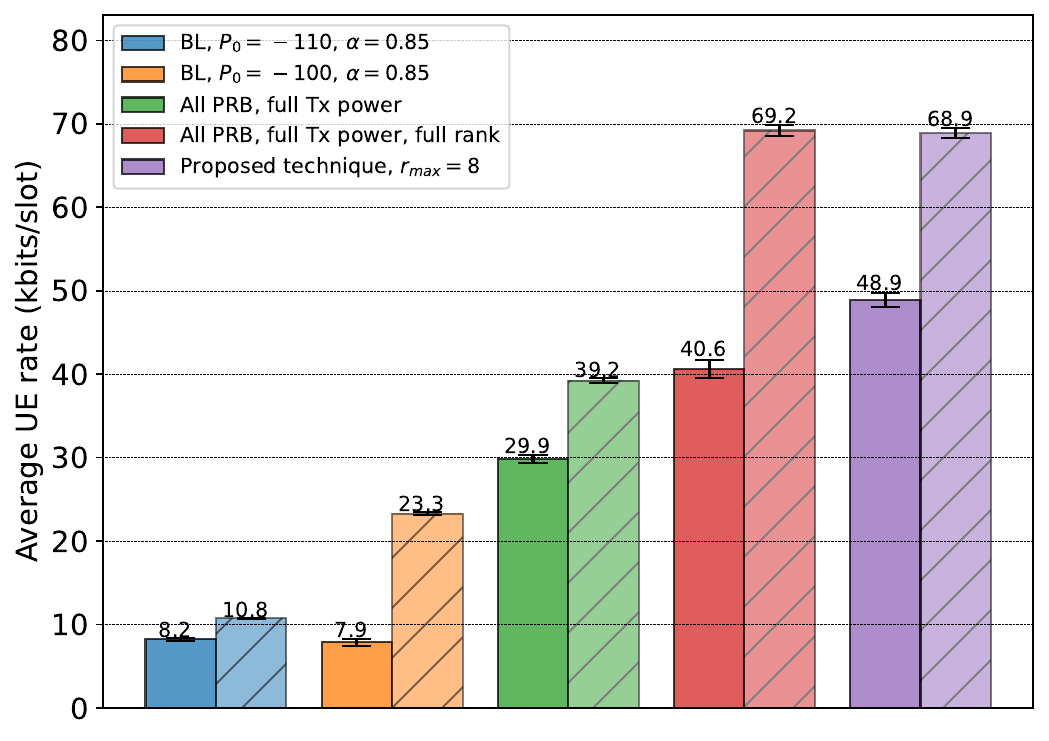}
      \caption{UL: Plots (with $90\%$ confidence interval) of the \gls{UE} \gls{GM} rates (plain bar) and \gls{AM} rates (hatched bars) in bits/slot for the isolated cell case.}
      \label{fig:ul_rate_isolated_cell}
\end{figure} 

\begin{figure*}[!t]
      \centering
      \includegraphics[scale = 0.35]{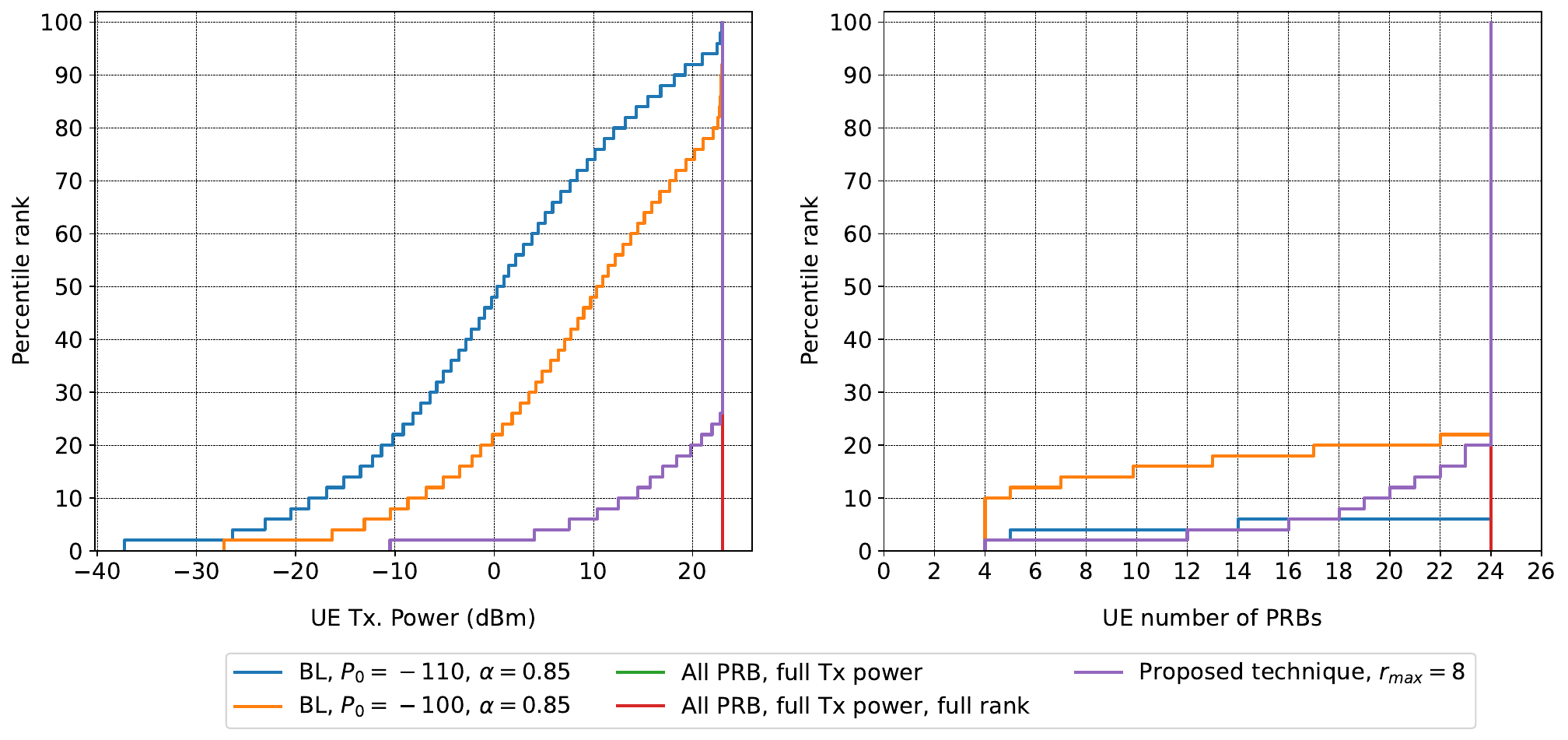}
      \caption{UL: Percentile ranks of the \gls{UE} transmit power (left) and the \gls{PRB} allocation for the isolated cell case. }
      \label{fig:ul_txp_prb_isolated_cell}
\end{figure*}

\begin{figure}[!t]
      \centering
      \includegraphics[width=3.4in]{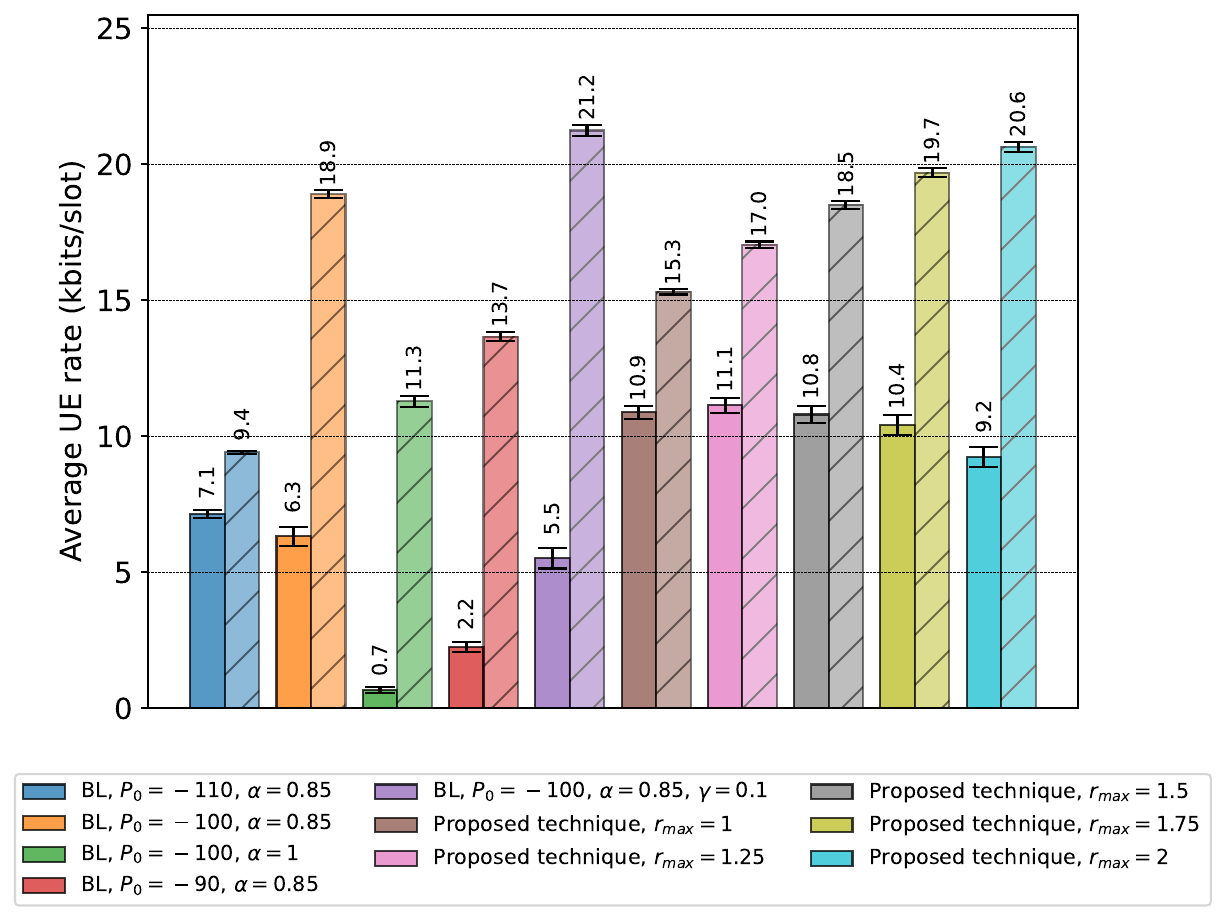}
      \caption{UL: Plots (with $90\%$ confidence interval) of the \gls{UE} \gls{GM} rates (plain bar) and \gls{AM} rates (hatched bars) in bits/slot with \gls{ICI}.}
      \label{fig:ul_rate_interf_cell}
\end{figure}

\begin{figure*}[!t]
      \centering
      \includegraphics[scale = 0.38]{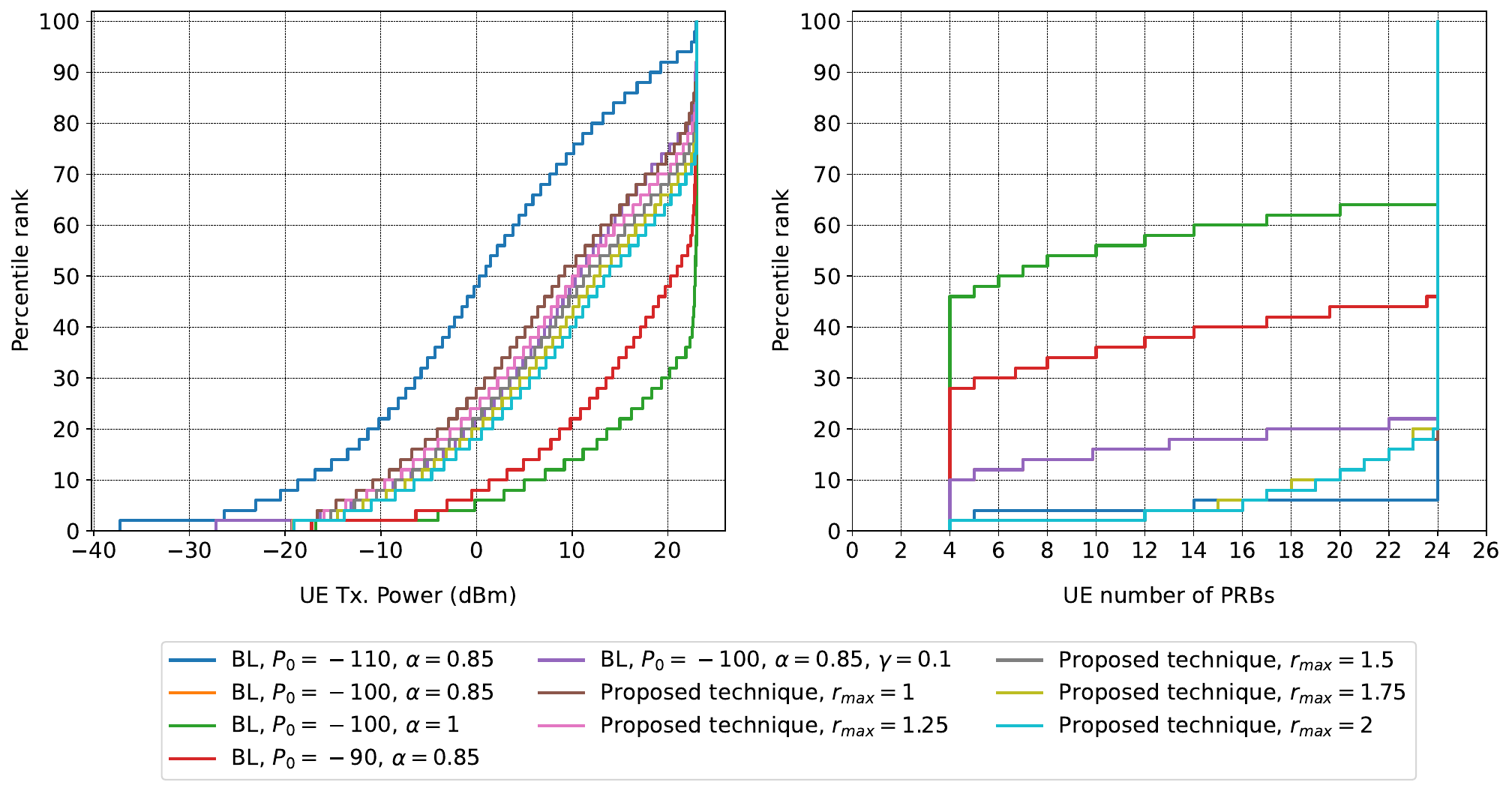}
      \caption{UL: Percentile ranks of the \gls{UE} transmit power (left) and the \gls{PRB} allocation (right) for the case with \gls{ICI}. }
      \label{fig:ul_txp_prb_interf_cell}
\end{figure*}

\begin{figure*}[!t]
      \centering
      \includegraphics[scale = 0.38]{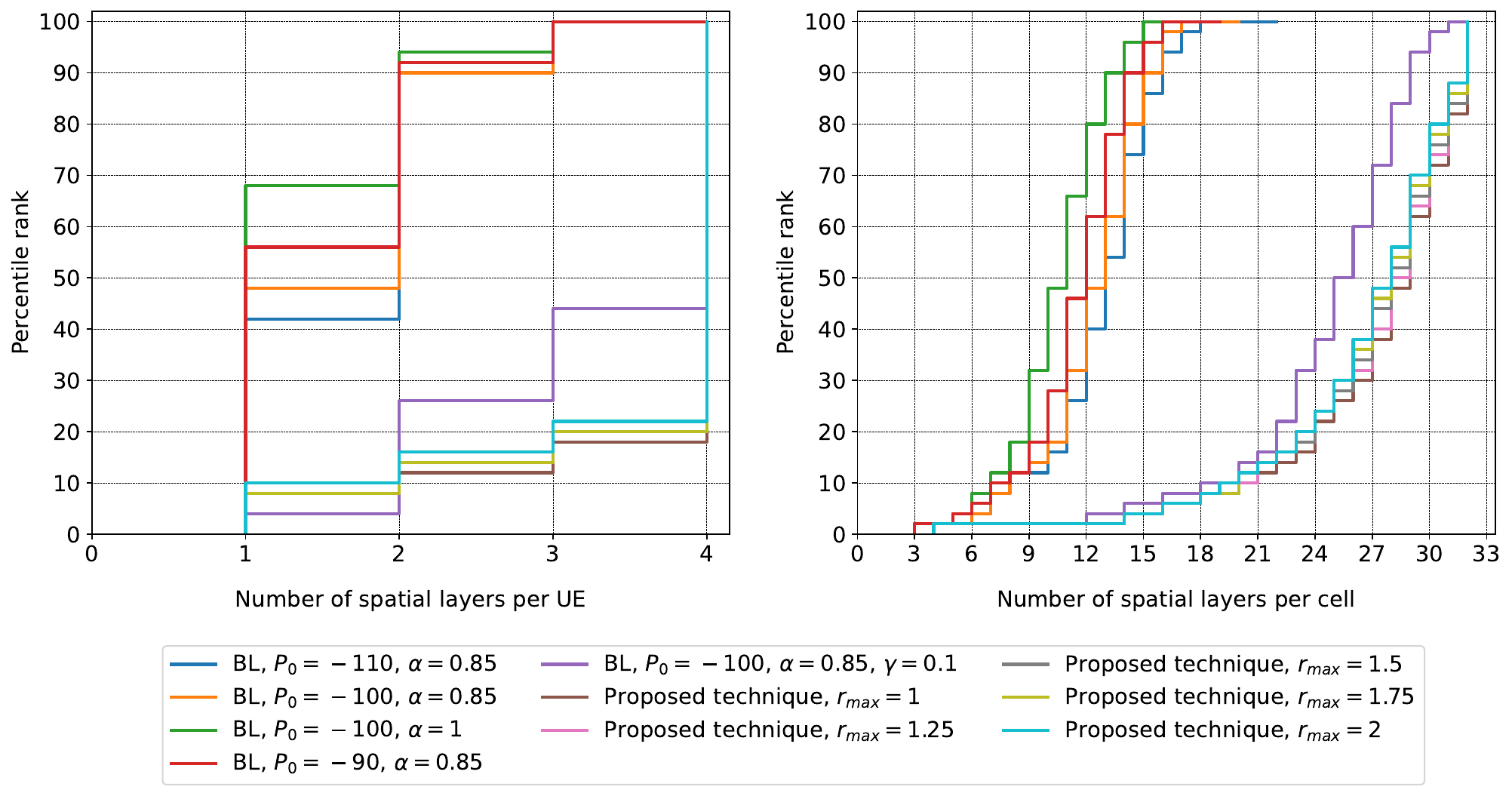}
      \caption{UL: Percentile ranks of the number of data layers per \gls{UE} and the number of spatial layers per \gls{BS} for the case with \gls{ICI}. }
      \label{fig:ul_layers_interf_cell}
\end{figure*}

\begin{figure}[!t]
      \centering
      \includegraphics[width=3.4in]{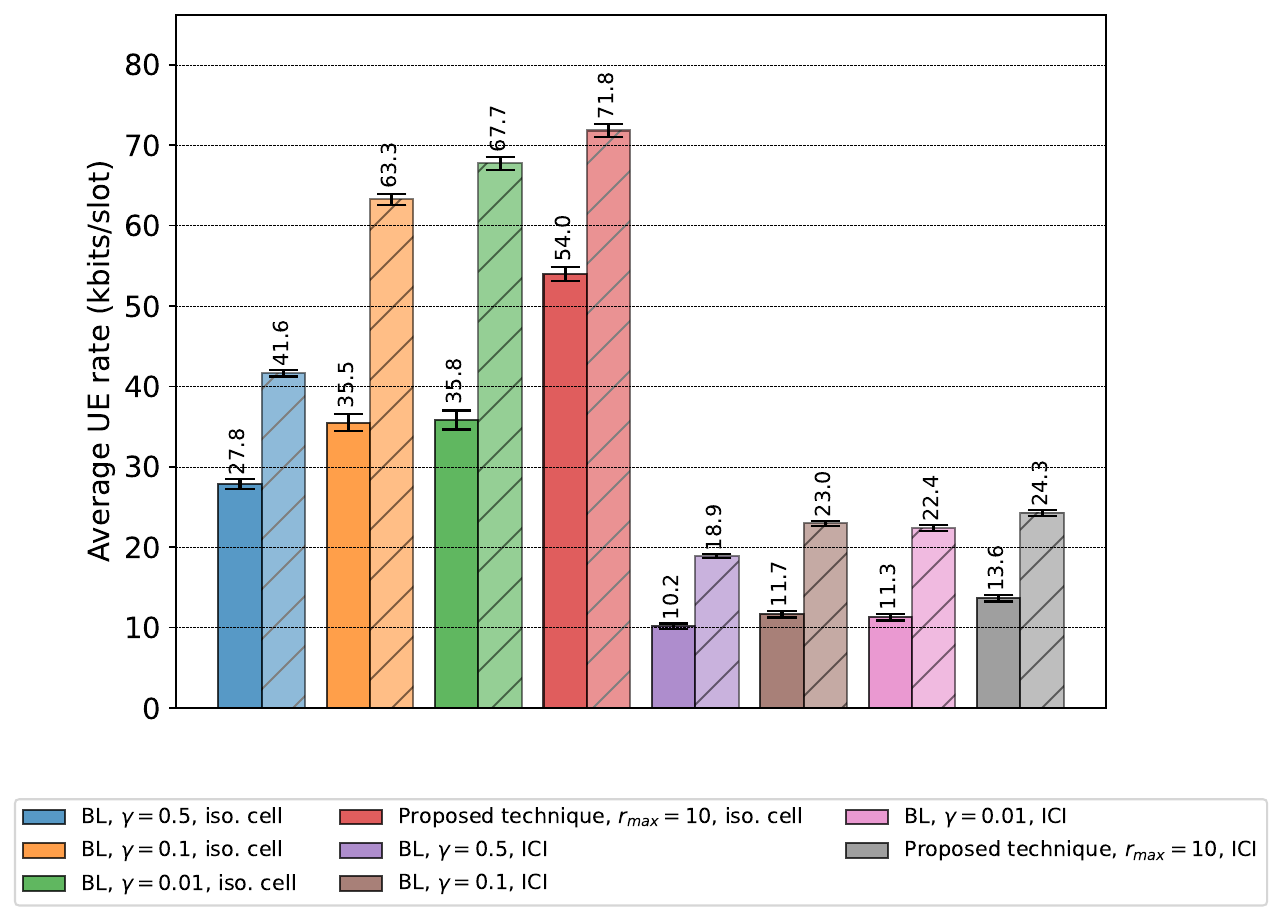}
      \caption{DL: Plots (with $90\%$ confidence interval) of the \gls{UE} \gls{GM} rates (plain bar) and \gls{AM} rates (hatched bars) in bits/slot.}
      \label{fig:dl_rate_iso_ICI_cell}
\end{figure}

\begin{figure*}[!t]
      \centering
      \includegraphics[scale = 0.4]{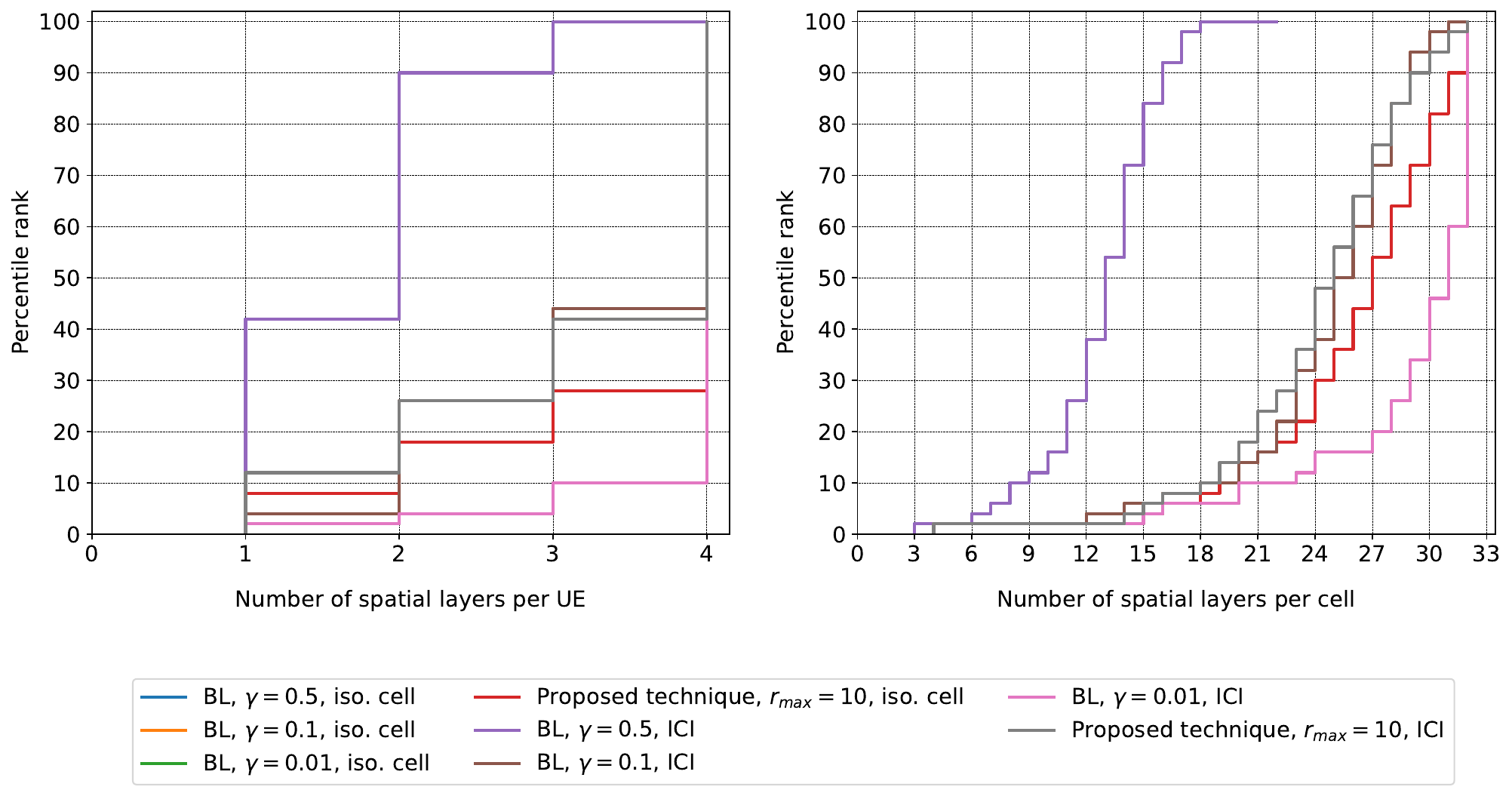}
      \caption{DL: Percentile ranks of the number of data layers per \gls{UE} and the number of spatial layers per \gls{BS}. }
      \label{fig:dl_layers_iso_ICI_cell}
\end{figure*}

\begin{figure*}[!t]
      \centering
      \includegraphics[scale = 0.4]{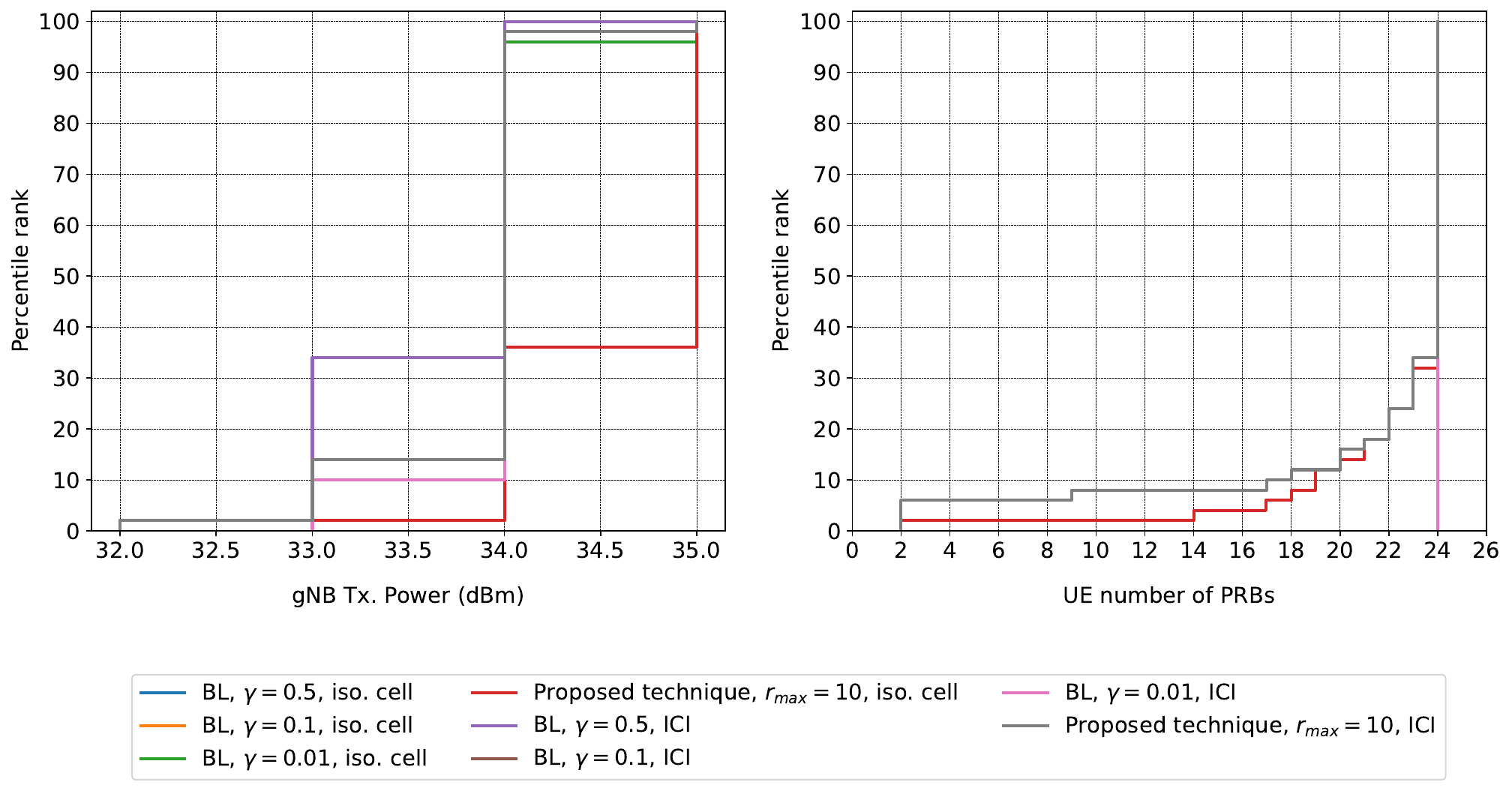}
      \caption{DL: Percentile ranks of the \gls{BS} transmit power (left) and the \gls{UE} \gls{PRB} allocation (right). }
      \label{fig:dl_txp_prb_iso_ICI_cell}
\end{figure*}

We consider the following setup (with a summary of the parameters in Table~\ref{tab:parameters}) for performing multi-cell, multi-link-level simulations, the code for which was written in NumPy (for channel generation) and TensorFlow. We consider 7 sites with 3 cells per site, leading to a total of $21$ cells arranged in a hexagonal grid with wraparound (which essentially means that each cell sees an \gls{ICI} pattern similar to that of the central cell). Each cell serves only at most $8$ \glspl{UE} per slot in \gls{MU-MIMO} mode. We consider the following two scenarios:
\begin{enumerate}
    \item All the cells are isolated which means that there is no \gls{ICI}. This corresponds to the case $\Rm_{f,t} = \sigma^2 \Id$ in \eqref{eq:ul_NW_model}, where $\sigma^2$ is the thermal noise variance.
    \item Cells with \gls{ICI} from the adjacent cells whose statistics vary from slot to slot depending on the paired \glspl{UE} in the adjacent cells. In this case, there is no way of obtaining an accurate enough estimate of $\Rm_{g}$ (introduced in Section~\ref{subsec:ul_problem}) at the \gls{BS} in order to accurately estimate the correct \gls{MCS} for all the co-scheduled \glspl{UE} unless the \gls{ICI} is lower than or comparable to the thermal noise level.
\end{enumerate} 
 For the uplink, we use QPSK, 16-QAM, 64-QAM, and 256-QAM while for the downlink, we also use 1024-QAM. Explicit  data transmission, receiver equalization, and channel coding/decoding using \gls{5GNR} \gls{LDPC} codes are performed. We assume the availability of perfect \gls{CSI} but no \gls{OLLA}~\cite{la_2020} or \gls{HARQ} whose inclusion, in practice, can be expected to make up for ill-effects of imperfect \gls{CSI}. Therefore, \gls{MCS} selection is done as detailed in~\cite{kps2023} while taking $\Rm_{g}$ to be $\sigma^2 \Id$ in \eqref{eq:ul_NW_model}, where $\sigma^2$ is the thermal noise variance per \gls{PRB}.

 \begin{table}[htbp]
      \begin{center}
        \caption{A list of simulation parameters.}
        \label{tab:parameters}
          \begin{tabular}{|c|c|}
          \hline
              {\bf Parameter}      & {\bf Value}                         \\ \hline \hline
              $N_{UE}$  & $8$ (maximum)   \\ \hline
              $n_U^{(i)}$   & $4$ ($2$H$\times 1$V$\times2$P), $\forall i$  \\ \hline
                    $n_B$  & $128$ ($4$H$\times 16$V$\times2$P) \\ \hline
              Number of cells & \\ in the network  & $21$ \\ \hline
                    Total number of \glspl{UE} & \\ in the network & $210$ \\ \hline
              Inter-site distance   & \SI{500}{\metre}   \\ \hline
              Channel model  & 38.901 Urban Macro (UMa) NLoS  \\ \hline
              Carrier frequency  & \SI{3.5}{\GHz}   \\ \hline              
              \gls{OFDM} subcarrier spacing   & \SI{30}{\kHz} \\ \hline	
               $T$& $14$ \\ \hline
              Slot duration &\SI{0.5}{\ms} \\ \hline
              Bandwidth & \SI{8.64}{\MHz} ($24$ \glspl{PRB}) \\ \hline
              $n_{RBG}$ & $24$ ($1$ \gls{PRB} per \gls{RBG}) \\ \hline
              $n_{RBG,min}$ & $4$ (uplink), $2$ (downlink) \\ \hline
              $P_{B,max}$  & \SI{36}{\dBm} for \SI{8.64}{MHz}  \\ \hline
              \gls{UE} speed   &\SI{3}{\kmph}  \\ \hline
               $P_{U,max}$  & \SI{23}{\dBm}  for \SI{8.64}{MHz}  \\ \hline
                  \gls{UE} \gls{OLPC} parameters &  $P_0$ (dBm) $\in \{ -85, -90, -100, -110\}$, \\
                   $(P_0, \alpha)$ &  $\alpha \in \{ 0.85, 1\}$   \\ \hline
             Channel coding & \gls{5GNR}  \gls{LDPC}~\cite[Section~5.3.2]{3GPP_dci_2020} \\ &  with rate-matching \\ \hline
             Channel estimation & Perfect \gls{CSI} \\ \hline
             \gls{MCS} & Table 5.1.3.1-2 (uplink), \\ & Table 5.1.3.1-4 (downlink) of~\cite{3GPP_MCS_table_2020} \\ \hline
             Scheduler & Plain Round Robin \\ \hline
              \Gls{HARQ} & No \\ \hline
              Number of independent & \\ \gls{UE} drops & 10 \\ \hline
              Number of slots per drop & 100 \\ \hline
              Receiver equalizer & \gls{LMMSE} \\ \hline
              Downlink precoder & Block diagonalization \gls{ZF}~\cite{stankovic08} \\ \hline
              CSI aquistion frequency & Once every $20$ slots  \\ \hline
              Number of iterations & \\ of IPOPT & $10$ \\ \hline
          \end{tabular}
      \end{center}    
  \end{table}

\subsection{Uplink Results} \label{subsec:uplink_sim}
\noindent  {\bf Baseline power control}: The total transmit power $P_i$ (dBm) for \gls{UE} $i$ is obtained by setting the \gls{OLPC} parameters $(P_0,\alpha)$ in the following~\cite[Section~7]{3GPP_UL_control_2021}: 
\begin{equation}\label{eq:OLPC}
	P_i = \min\{ P_{U,max}, P_0 + 10\logt{n_{PRB,i}} + \alpha PL_i \}
\end{equation}
where $ P_{U, max} = \SI{23}{\dBm}$ for the bandwidth considered, $n_{PRB,i}\in [4, 24]$ is the total number of \glspl{PRB} assigned to \gls{UE} $i$, $PL_i$ is the pathloss estimate (based on the measured channel gains on the downlink), $P_0$ (dBm) is the expected received power per \gls{PRB} under full pathloss compensation, and $\alpha \in [0,1]$ is the fractional pathloss compensation factor. 

\noindent  {\bf Baseline \gls{RBG} allocation}: We assume that all the \glspl{RBG} are allocated to the \gls{UE} unless power-limited. In the latter case, we assign a number of \glspl{RBG} (subject to a minimum of $4$) over which the \gls{UE} has sufficient power to transmit, choosing the available \glspl{RBG} with the strongest channel gains for that \gls{UE}. 

\noindent {\bf Baseline \gls{UE} rank selection}: Let the channel covariance (at the \gls{UE}) for UE $i$ be $\Rm_{h,UE}^{(i)} \triangleq \Ebb_{f,t} \LSB \Hm_{i,f,t}^H \Hm_{i,f,t}  \RSB \in  \Cbb^{n_U^{(i)} \times n_U^{(i)}}$ with ordered eigenvalues $\mu_1 \geq \mu_2 \geq \cdots \geq \mu_{n_U^{(i)}}$. The \gls{UE} rank $n_l^{(i)}$ for the baseline scheme is taken to be $n_l^{(i)} = \max \LP n \in \LP 1,\cdots, n_U^{(i)} \RP \Big \vert \mu_n/\mu_1 \geq  \gamma  \RP$ where $ \gamma \in (0,1]$  is a predefined threshold and taken to be $0.5$, $0.1$, or $0.01$ in this study. Further, $n_l^{(i)}=1$ if $n_{RBG,i} = n_{RBG,min}$.

Apart from these baseline schemes, we also consider the case where all the \glspl{PRB} are allocated to all the \glspl{UE} with full transmit power and also with full rank ($n_l^{(i)} = n_U^{(i)}, \forall i$). For the proposed scheme, Algorithm \ref{alg:pseudocode} is executed but {\it a uniform transmit power across layers and \glspl{PRB} is applied by taking the mean of the obtained \gls{PRB}-wise and layer-wise powers for the \gls{UE} in context}.

Fig.~\ref{fig:ul_rate_isolated_cell} shows the comparison (``BL" indicates ``baseline") of the average \gls{GM} and arithmetic mean (AM) \gls{UE} rates, both computed per slot and then averaged across slots and independent \gls{UE} drops. While the proposed technique is not targeted at maximizing the \gls{AM} rates, they have been shown purely for illustration. Since there is no \gls{ICI}, it is obvious that aggressively using as much of the available transmit power as required is generally the best strategy. Therefore, we use $r_{min} = 0.23$ (corresponding to 4-QAM with the least coding rate in \gls{5GNR}) and $r_{max} = 8$ (corresponding to the upper bound on the rate per data symbol for 256-QAM) for the proposed technique (Section~\ref{subsec:ul_problem}). While the proposed technique not only achieves the best \gls{GM} \gls{UE} rates, it also does so while requiring lower \gls{UE} transmit powers than the naive, full transmit power scheme, as shown in Fig.~\ref{fig:ul_txp_prb_isolated_cell}.

For the case with \gls{ICI}, since the transmit powers of the served \glspl{UE} affect the reception in the neighbouring cells and the statistics of this \gls{ICI} varies slot-to-slot, the best performing techniques usually choose the transmit powers conservatively so as to minimize the \gls{ICI}. Otherwise, \gls{MCS} selection would be significantly affected even for the baseline schemes. So, we conservatively take $r_{max}$ to be between $1$ and $2$. We also consider $\gamma = 0.1$ for one of the baseline schemes so that a higher number of layers is selected for each \gls{UE}. Fig.~\ref{fig:ul_rate_interf_cell} highlights that the proposed scheme provides a significantly higher ($> 50\%$) \gls{GM} rates compared to the best baseline schemes while requiring a similar level of transmit powers per \gls{UE} (for most baseline schemes) as shown in Fig.~\ref{fig:ul_txp_prb_interf_cell}. The empirical \glspl{CDF} of the number of spatial layers per \gls{UE} and the total number of spatial layers per \gls{BS} are shown in Fig.~\ref{fig:ul_layers_interf_cell}. The trend clearly shows that while the \gls{AM} rates can be improved by increasing $r_{max}$ at the cost of \gls{GM} rates, the best balance is achieved for $r_{max}$ between $1.25$ and $1.75$. 

\subsection{Downlink Results} \label{subsec:downlink_sim}

\noindent  {\bf Baseline power allocation}: Let $\wv_{g,row}^{(k)}$ denote the $k^{th}$ row of $\Wm_g$, 
$k=1,\cdots, n_B$, $g=1,\cdots, n_{RBG}$, where $\Wm_g$ is the equivalent of $\Wm_{f,t}$ mentioned in Section~\ref{subsec:DL} at a \gls{RBG} level. The columns of $\Wm_g$ are normalized to unity. Then, the precoder is taken to be $\Wm_g \leftarrow \sqrt{P} \Wm_g$ with $P=n_{RBG}P_{ant}/max_{k=1,\cdots,n_B}\LP \sum_{g \in \Gc}n_F\Vert w_{g,row}^{(k)}\Vert^2 \RP $, where $n_F/n_{RBG} = 12$, the number of subcarriers per \gls{RBG}. 

\noindent  {\bf Baseline \gls{RBG} allocation}: For lack of a better alternative baseline scheme, we assume that all the \glspl{RBG} are allocated to all the \glspl{UE}. Note that the techniques highlighted in~\cite{Faroq_leasch},~\cite{AI6GMAC} require reinforcement-learning and multiple PRB-looping which are out of scope of this paper.

\noindent {\bf Baseline \gls{UE} rank selection}: This is done in the same way as for the uplink.

For the purpose of the estimating the covariance of the interference-plus-noise at the \gls{UE} (Section~\ref{subsec:dl_problem}), we assume that the \gls{UE} $i$ measures the interference-plus-noise power $\sigma^2_{i}$ and feeds this scalar back to the serving \gls{BS}. So, the \gls{BS} approximates the covariance of this interference-plus-noise as $\sigma^2_{i} \Id$. Fig.~\ref{fig:dl_rate_iso_ICI_cell} shows the plots of the \gls{GM} and the \gls{AM} \gls{UE} rates for the baseline scheme (with different values of $\gamma$ for layer selection) and the proposed scheme. The empirical \glspl{CDF} of the number of spatial layers per \gls{UE} and the total number of spatial layers per \gls{BS} are shown in Fig.~\ref{fig:dl_layers_iso_ICI_cell} while the \glspl{CDF} of the \gls{BS} transmit power and the \gls{PRB} allocation are shown in Fig.~\ref{fig:dl_txp_prb_iso_ICI_cell}. These plots reveal that for the isolated cell scenario, there is a significant improvement in the \gls{GM} rates ($>50\%$ compared to the best baseline scheme) while for the case with \gls{ICI}, there is around $18\%$ improvement.

%% file: tex/concluding_remarks.tex
\section{Concluding Remarks}
\label{sec:conc_remarks}

In this work, we proposed a unified framework for jointly performing frequency resource allocation, power allocation/control, and \gls{UE} rank selection in \gls{MU-MIMO} systems equipped with linear transceivers. This framework encompasses both uplink and downlink data transmission scenarios. We presented a computationally-efficient algorithm to achieve a near-optimal solution to this problem. Through extensive simulations, we demonstrated that the proposed technique can significantly improve the geometric-mean \gls{UE} rates compared to existing baseline schemes. While the proposed approach requires the application of interior-point methods in every time slot, we posit the possibility of leveraging supervised or reinforcement learning techniques to guide the solution process. By training these algorithms with expert policies based on the proposed technique, one can potentially eliminate the need for on-the-fly iterative optimization. This avenue presents a promising direction for future research endeavors.

%% file: main.bbl
\begin{thebibliography}{10}
\providecommand{\url}[1]{#1}
\csname url@samestyle\endcsname
\providecommand{\newblock}{\relax}
\providecommand{\bibinfo}[2]{#2}
\providecommand{\BIBentrySTDinterwordspacing}{\spaceskip=0pt\relax}
\providecommand{\BIBentryALTinterwordstretchfactor}{4}
\providecommand{\BIBentryALTinterwordspacing}{\spaceskip=\fontdimen2\font plus
\BIBentryALTinterwordstretchfactor\fontdimen3\font minus
  \fontdimen4\font\relax}
\providecommand{\BIBforeignlanguage}[2]{{%
\expandafter\ifx\csname l@#1\endcsname\relax
\typeout{** WARNING: IEEEtran.bst: No hyphenation pattern has been}%
\typeout{** loaded for the language `#1'. Using the pattern for}%
\typeout{** the default language instead.}%
\else
\language=\csname l@#1\endcsname
\fi
#2}}
\providecommand{\BIBdecl}{\relax}
\BIBdecl

\bibitem{stankovic08}
V.~Stankovic and M.~Haardt, ``Generalized design of multi-user mimo precoding
  matrices,'' \emph{IEEE Transactions on Wireless Communications}, vol.~7,
  no.~3, pp. 953--961, 2008.

\bibitem{chou_adaptive_rank}
W.-H. Chou, W.-C. Pao, C.-C. Tsai, T.-Y. Yeh, and J.-Y. Pan, ``An adaptive rank
  selection method in 3gpp 5g nr systems,'' in \emph{2021 Asia-Pacific Signal
  and Information Processing Association Annual Summit and Conference (APSIPA
  ASC)}, 2021, pp. 1912--1916.

\bibitem{maggi21}
L.~Maggi, A.~Valcarce, and J.~Hoydis, ``Bayesian optimization for radio
  resource management: Open loop power control,'' \emph{IEEE Journal on
  Selected Areas in Communications}, vol.~39, no.~7, pp. 1858--1871, 2021.

\bibitem{Xuejun_adaptive_resource_allocation}
X.~Sha, J.~Sun, T.~Wu, and L.~Liu, ``Adaptive resource allocation based packet
  scheduling for lte uplink,'' in \emph{2012 IEEE 11th International Conference
  on Signal Processing}, vol.~2, 2012, pp. 1473--1476.

\bibitem{zhao23}
X.~Zhao, S.~Lu, Q.~Shi, and Z.-Q. Luo, ``Rethinking wmmse: Can its complexity
  scale linearly with the number of bs antennas?'' \emph{IEEE Transactions on
  Signal Processing}, vol.~71, pp. 433--446, 2023.

\bibitem{Faroq_leasch}
F.~Al-Tam, N.~Correia, and J.~Rodriguez, ``Learn to schedule (leasch): A deep
  reinforcement learning approach for radio resource scheduling in the 5g mac
  layer,'' \emph{IEEE Access}, vol.~8, pp. 108\,088--108\,101, 2020.

\bibitem{NokiaDeepScheduler2024}
P.~Kela, B.~Liu, A.~Valcarce \emph{et~al.}, ``Towards a deep scheduler for
  allocating radio resources,'' 2024, in preparation.

\bibitem{3GPP_MCS_table_2020}
\BIBentryALTinterwordspacing
3GPP, ``{NR; Physical layer procedures for data},'' {3rd Generation Partnership
  Project (3GPP)}, Technical Specification (TS) 38.214, 03 2024, version
  18.2.0. [Online]. Available:
  \url{https://portal.3gpp.org/desktopmodules/Specifications/SpecificationDetails.aspx?specificationId=3216}
\BIBentrySTDinterwordspacing

\bibitem{Tse2005}
D.~Tse and P.~Viswanath, \emph{{Fundamentals of Wireless Communication}}.\hskip
  1em plus 0.5em minus 0.4em\relax USA: Cambridge University Press, 2005.

\bibitem{Caire1998}
G.~{Caire}, G.~{Taricco}, and E.~{Biglieri}, ``{Bit-Interleaved Coded
  Modulation},'' \emph{{IEEE Trans. Inf. Theory}}, vol.~44, no.~3, pp.
  927--946, May 1998.

\bibitem{Uchida_alpha_fair}
M.~Uchida and J.~Kurose, ``An information-theoretic characterization of
  weighted alpha-proportional fairness,'' in \emph{IEEE INFOCOM 2009}, 2009,
  pp. 1053--1061.

\bibitem{Kelly1998RateCF}
\BIBentryALTinterwordspacing
F.~Kelly and A.~Maulloo, ``Rate control for communication networks: shadow
  prices, proportional fairness and stability,'' \emph{Journal of the
  Operational Research Society}, vol.~49, pp. 237--252, 1998. [Online].
  Available: \url{https://api.semanticscholar.org/CorpusID:2876114}
\BIBentrySTDinterwordspacing

\bibitem{dikin1967iterative}
I.~I. Dikin, ``Iterative solution of problems of linear and quadratic
  programming,'' \emph{Dokl. Akad. Nauk {SSSR}}, vol. 174, pp. 747--748, 1967.

\bibitem{Wachter2006}
\BIBentryALTinterwordspacing
A.~W{\"a}chter and L.~T. Biegler, ``On the implementation of an interior-point
  filter line-search algorithm for large-scale nonlinear programming,''
  \emph{Mathematical Programming}, vol. 106, no.~1, pp. 25--57, Mar 2006.
  [Online]. Available: \url{https://doi.org/10.1007/s10107-004-0559-y}
\BIBentrySTDinterwordspacing

\bibitem{3GPP_dci_2020}
\BIBentryALTinterwordspacing
3GPP, ``{NR; Multiplexing and channel coding},'' {3rd Generation Partnership
  Project (3GPP)}, Technical Specification (TS) 38.212, 12 2020, version
  16.4.0. [Online]. Available:
  \url{https://portal.3gpp.org/desktopmodules/Specifications/SpecificationDetails.aspx?specificationId=3214}
\BIBentrySTDinterwordspacing

\bibitem{wright_IP_methods}
\BIBentryALTinterwordspacing
S.~J. Wright, \emph{Primal-Dual Interior-Point Methods}.\hskip 1em plus 0.5em
  minus 0.4em\relax Society for Industrial and Applied Mathematics, 1997.
  [Online]. Available:
  \url{https://epubs.siam.org/doi/abs/10.1137/1.9781611971453}
\BIBentrySTDinterwordspacing

\bibitem{la_2020}
S.~Sun, S.~Moon, and J.-K. Fwu, ``{Practical Link Adaptation Algorithm With
  Power Density Offsets for 5{G} Uplink Channels},'' \emph{IEEE Wireless
  Communications Letters}, vol.~9, no.~6, pp. 851--855, 2020.

\bibitem{kps2023}
K.~P. Srinath and J.~Hoydis, ``Bit-metric decoding rate in multi-user mimo
  systems: Applications,'' \emph{IEEE Transactions on Wireless Communications},
  vol.~22, no.~6, pp. 3968--3981, 2023.

\bibitem{3GPP_UL_control_2021}
\BIBentryALTinterwordspacing
3GPP, ``{NR; Physical layer procedures for control},'' {3rd Generation
  Partnership Project (3GPP)}, Technical Specification (TS) 38.213, 12 2021,
  version 17.0.0. [Online]. Available:
  \url{https://portal.3gpp.org/desktopmodules/Specifications/SpecificationDetails.aspx?specificationId=3215}
\BIBentrySTDinterwordspacing

\bibitem{AI6GMAC}
A.~Valcarce, P.~Kela, S.~Mandelli, and H.~Viswanathan, ``The role of ai in 6g
  mac,'' in \emph{2024 Joint European Conference on Networks and Communications
  \& 6G Summit (EuCNC/6G Summit)}, 2024.

\end{thebibliography}
